\documentclass[12pt]{article}
\usepackage{amsmath,amsfonts,sint,epsfig,cite,graphics}
\usepackage{amssymb,bbm}
\usepackage{mathrsfs} 
\usepackage{amsbsy,color}
\usepackage{latexsym}
\usepackage{graphicx}
\usepackage{psfrag}
\usepackage{multirow}
\usepackage{booktabs}
\usepackage{url}
\usepackage[T1]{fontenc} 

\newcommand{\be}{\begin{equation}}
\newcommand{\ee}{\end{equation}}
\newcommand{\bea}{\begin{eqnarray}}
\newcommand{\eea}{\end{eqnarray}}
\newcommand{\bes}{\begin{eqnarray}}
\newcommand{\ees}{\end{eqnarray}}
\newcommand{\ba}{\begin{array}}
\newcommand{\ea}{\end{array}}

\newcommand{\eq}[1]{eq.~(\ref{#1})}

\newcommand{\eqs}[1]{eqs.~(\ref{#1})}

\newcommand{\fig}[1]{Fig.~\ref{#1}}
\newcommand{\Fig}[1]{Figure~\ref{#1}}
\newcommand{\sect}[1]{Sect.~\ref{#1}}

\newcommand{\app}[1]{App.~\ref{#1}}

\newcommand{\tab}[1]{Table~\ref{#1}}
\newcommand{\Tab}[1]{Table~\ref{#1}}
\newcommand{\Tabs}[1]{Tables~\ref{#1}}
\newcommand{\Ref}[1]{Ref.~\cite{#1}}

\newcommand{\tauint}{\tau_\mathrm{int}}
\newcommand{\psibar}{\overline \psi}

\newcommand{\rmO}{\mathrm{O}}

\newcommand{\mr}{m_\mathrm{R}}
\newcommand{\msea}{m_\mathrm{sea}}
\newcommand{\fk}{f_\mathrm{K}}
\newcommand{\fklat}{{F}_\mathrm{K}}
\newcommand{\fklatphys}{F_{\mathrm{K,phys}}}
\newcommand{\fkphys}{f_{\mathrm{K,phys}}}
\newcommand{\yk}{y_{\mathrm{K}}}
\newcommand{\ypi}{y_{\pi}}
\newcommand{\mksqlat}{M^2_{\mathrm{K}}}
\newcommand{\mulat}{\mu}
\newcommand{\mustrange}{\mu_{\strange}}
\def\nf{N_{\rm f}}
\newcommand{\mk}{m_{\rm K}}
\newcommand{\mkphys}{m_{\rm K,phys}}
\newcommand{\strange}{{\rm s}}
\newcommand{\light}{{\rm light}}

\newcommand{\mstrange}{m_{\strange}}

\newcommand{\Mstrange}{M_{\strange}}
\newcommand{\Mlight}{M_\light}
\newcommand{\mlight}{m_\light}
\newcommand{\mpi}{m_{\pi}}
\newcommand{\mpiphys}{m_{\pi,\mathrm{phys}}}
\newcommand{\fpi}{f_{\pi}}
\newcommand{\fpiphys}{f_{\pi,\rm phys}}
\def\mbar{\overline{m}}
\def\gbar{\overline{g}}
\newcommand{\msbar}{{\rm \overline{MS\kern-0.05em}\kern0.05em}}

\def\ca{c_{\rm A}}
\def\za{Z_{\rm A}}
\def\zp{Z_{\rm P}}
\def\ba{b_{\rm A}}
\def\bp{b_{\rm P}}
\def\babar{\bar b_{\rm A}}
\def\bpbar{\bar b_{\rm P}}
\def\batil{\tilde b_{\rm A}}
\def\bptil{\tilde b_{\rm P}}

\def\MeV{{\rm MeV}}
\newcommand{\lmax}{L_{\rm max}}

\def\R0c{R_{0{\rm c}}}
\def\r0ref{r_{0{\rm ref}}}

\begin{document}

\begin{titlepage}

\begin{flushright}
\small{
CERN-PH-TH-2012-119\\
DESY 12-078 \\
HU-EP-12/15\\
SFB/CPP-12-24\\
WUB/12-11
}
\end{flushright}

\begin{center}
{\Large\bf
The strange quark mass and Lambda parameter of two flavor QCD
}
\end{center}
\vskip 0.35cm
\vbox{
\centerline{
\epsfxsize=2.8 true cm
\epsfbox{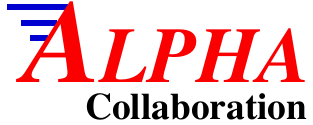}}
}
\vskip 0.1cm
\begin{center}
{
Patrick Fritzsch,$^{\scriptscriptstyle a}$
Francesco Knechtli,$^{\scriptscriptstyle b}$
Bj\"orn Leder,$^{\scriptscriptstyle b}$
Marina Marinkovic,$^{\scriptscriptstyle a}$\\
Stefan Schaefer,$^{\scriptscriptstyle c}$
Rainer Sommer$^{\scriptscriptstyle d}$ and
Francesco Virotta$^{\scriptscriptstyle d}$
}
\vskip 0.5cm
{
\vskip 2.0ex
$^{\scriptstyle a}$
        Humboldt Universit\"at zu Berlin, Institut f\"ur Physik, \\
	Newtonstr.~15, 12489~Berlin, Germany
\vskip 2.0ex
$^{\scriptstyle b}$
        Bergische Universit\"at Wuppertal, \\
	Fachbereich C -- Mathematik und Naturwissenschaften, \\
	Gaussstr. 20, 42119~Wuppertal, Germany
\vskip 2.0ex
$^{\scriptstyle c}$
        CERN, Physics Department, 1211~Geneva~23, Switzerland 
\vskip 2.0ex
$^{\scriptstyle d}$
        NIC, DESY, Platanenallee 6, 15738~Zeuthen,  Germany
\vskip 2.0ex
}
\vskip 0.5cm
{\bf Abstract}
\vskip 0.1ex
\end{center}

We complete the non-perturbative calculations of the strange quark
mass and the Lambda parameter in two flavor QCD by the ALPHA
collaboration. The missing lattice scale is determined via the 
kaon decay constant, for whose chiral extrapolation 
complementary strategies are compared. We also give a value for 
the scale $r_0$ in physical units as well as an improved
determination of the renormalization constant $\za$.


\vskip 2.0ex

\vfill
\eject

\end{titlepage}

\section{Introduction}

The parameters of the standard model of particle physics have to be
determined by matching  theory to experimental data.  For two flavor QCD
we present results for two of these parameters, the scale parameter
$\Lambda_\mathrm{QCD}$ and the mass of the strange quark.

These results are the outcome of a long project by the ALPHA
collaboration. The general strategy using the Schr\"odinger functional
to define the coupling constant has been laid out in
Refs.~\cite{alpha:sigma,SF:LNWW}, and results for the $\Lambda$
parameter in pure gauge theory have been published in \cite{alpha:SU3}.
In the two flavor theory for $\Lambda$\cite{alpha:letter,alpha:nf2} and
the strange quark mass\cite{mbar:nf2}, however, the determination of the
lattice scale, which allows for their conversion to physical units, has
been lacking. This determination is the subject of the present paper.

In lattice computations, the physical mass scale is set by picking one
dimensionful observable and identifying its value at the point where the
quark mass ratios correspond to the physical situation with the
experimental input.  If we have done the calculation with all physical
effects  taken into account (and if the theory is correct), it does not
matter, which observable we take.  Here, however, we restrict ourselves
to QCD with two dynamical flavors of light quarks leading to a
systematic uncertainty which is hard to determine. In our computation,
we use the kaon decay constant to set the scale.  Over the
pion decay constant it has the advantage of a chiral
extrapolation which is milder and therefore better under
control.\footnote{This property depends on how one actually
approaches the physical point.  It is in particular true
for the strategy 1 which we introduce below.}
However, we need a quenched strange quark. Also the
mass of the Omega baryon is popular to set the scale
and first results indicate that similar numbers are
obtained from this observable\cite{georg11}.

In previous publications the results have been converted to physical
units using the scale parameter $r_0$\cite{pot:r0}, defined via the
force between static quarks. The conversion relied on  measurements of
$r_0/a$ by QCDSF\cite{QCDSF:nf2mstrange} and the assumption that
$r_0=0.5$\,fm.  Below, we present our own results for $r_0/a$, which differ
substantially from the previous values and lead to an update in $r_0\Lambda$.

The paper is organized as follows. The lattice action,  an overview
of the ensembles and details of the error analysis are given in
\sect{s:2}, followed by the definition of the hadronic observables in
\sect{s:3} and results for the scale parameter $r_0$ in \sect{s:r0}. 
The strategies for the chiral extrapolation are discussed in \sect{s:5}
leading to the scale determination from the kaon decay constant. The results
for the Lambda parameter and the strange quark mass are contained in 
Sects. \ref{s:6} and \ref{s:7}, respectively.

The appendices contain updates of many quantities, whose analysis has
been subject of previous publications. The renormalization constants
$\za$ and $\zp$ are discussed in \app{a:ZA} and \app{a:ZP}, the hadronic
scale $L_1$ of the Schr\"odinger functional calculations is subject
of \app{a:l1}, followed by a determination of the critical mass of 
the improved Wilson fermions and the singlet renormalization factor 
in \app{a:mcrit}.

\section{Lattice parameters and simulation algorithms\label{s:2}}

In this computation we use the Wilson plaquette gauge action  for
the gluon fields together with two degenerate flavors of
$\rmO(a)$ improved Wilson fermions\cite{Wilson}. The action 
\be
S[U,\psibar,\psi]=
\beta \sum_p \mathrm{tr} \left [1-U(p)\right ]
+a^4 \sum_x \psibar(x)(D+m_0)\psi(x)
\ee
has three parameters: $g_0$, $m_0$ and $c_\mathrm{sw}$. 
The coupling constant $g_0$ is given by
 $\beta=6/g_0^2$. The fermions with bare mass $m_0$, usually substituted
by the hopping parameter $\kappa=(8+2 a m_0)^{-1}$, are implemented by the 
 lattice Dirac operator 
\be
D=\frac{1}{2}\{\gamma_\mu
( \nabla^*_\mu+\nabla_\mu)-a\nabla^*_\mu \nabla_\mu\}+
c_\mathrm{sw}  \frac{ia}{4}\sigma_{\mu\nu}\hat F_{\mu\nu}\, .
\ee
It includes the covariant forward and backward derivatives,
$\nabla_\mu$ and $\nabla_\mu^*$, and  the
Sheikholeslami--Wohlert\cite{impr:SW}
 improvement term involving the standard discretization $\hat
F_{\mu\nu}$ of the field strength tensor\cite{impr:pap1}.
Its coefficient $c_\mathrm{sw}$ has been determined
non-perturbatively\cite{impr:csw_nf2}.

We have generated ensembles at three values of $\beta=5.2$, 5.3 and 
5.5 which correspond roughly to lattice spacings of $a=0.076$\,fm,
$0.066$\,fm and $0.049$\,fm, respectively, with details given in Section~\ref{sec:6}.
The ensembles are listed in Table~\ref{tab:ens}. All lattices have 
size $(2L)\times L^3$ and the pion mass is always large enough such that
$m_\pi L\geq4$. We therefore expect finite size effects to be small.

\begin{table}
\small
\begin{center}
\begin{tabular}{@{\extracolsep{0.2cm}}cccccccc}
\toprule
id & $L/a$ & $\beta$ & $\kappa$  &  $\kappa_\strange$ & $R_0$ & $m_\pi$[MeV] & $m_\pi L$  \\
\midrule
A2  & $32$    &$5.2$    &$0.13565$& $0.135438(20)$ & $5.485(21)$&$630$ & $7.7$ \\ 
A3  &       &         &$0.13580$& $0.135346(20)$ & $5.674(32)$&$490$ & $6.0$ \\     
A4  &       &         &$0.13590$& $0.135285(20)$ & $5.808(34)$&$380$ & $4.7$ \\
A5  &       &         &$0.13594$& $0.135257(20)$ & $5.900(24)$&$330$ & $4.0$ \\
\midrule  
E4 & $32$  &$5.3$    &$0.13610$& $0.135836(17)$ &   ---      &$ 580$ & $6.2$ \\
E5 &       &         &$0.13625$& $0.135777(17)$ &   $6.747(59)$&$ 440$ & $4.7$   \\
F6 & $48$  &         &$0.13635$& $0.135741(17)$ &   $6.984(51)$&$ 310$ & $5.0$   \\
F7 &       &         &$0.13638$& $0.135730(17)$ &   $7.051(43)$&$ 270$ & $4.3$   \\
\midrule
N4  &$48$   &$5.5$    &$0.13650$& $0.136278(08)$ & $9.32(30)$&$550$     & $6.5$  \\
N5  &       &         &$0.13660$& $0.136262(08)$ & $9.31(26)$&$440$     & $5.2$  \\
N6  &       &         &$0.13667$& $0.136250(08)$ & $9.55(11)$&$340$     & $4.0$  \\
O7  &$64$   &         &$0.13671$& $0.136243(08)$ & $9.68(10)$&$270$     & $4.2$  \\
\bottomrule
\end{tabular}
\end{center}
\caption{\label{tab:ens}\footnotesize Overview of the ensembles used in this
study. We give the label, the spatial extent of the lattice,
$\beta=6/g_0^2$, the hopping parameter $\kappa$ of the sea quarks, 
the hopping parameter $\kappa_\strange$ of the strange quark,
the scale $R_0=r_0/a$, the mass of the 
sea pion $ m_\pi$ and the product $m_\pi L$,
which is always larger than or equal to $4$. All lattices have dimension
$T\times L^3$ with $T=2L$.}
\end{table}

\subsection{Simulation algorithms}

For most of the ensembles,  generated within the
CLS effort,\footnote{\url{https://twiki.cern.ch/twiki/bin/view/CLS/}}
the DD-HMC algorithm\cite{algo:L2,algo:L3} has been used as implemented
in the software package by M.~L\"uscher\cite{soft:DDHMC}. It is based on
a domain decomposition to separate the infrared from the ultraviolet
modes of the fermion determinant. A main feature is the locally
deflated, Schwarz preconditioned GCR
solver\cite{algo:L1,algo:coherence} which significantly  reduces 
the increase in computational cost as the quark mass is lowered.

The drawback of this algorithm is that due to the block decomposition
only a fraction of gauge links $R_\mathrm{active}$ is updated during 
a trajectory. In pure gauge theory the autocorrelation
times are inversely proportional to this fraction of active 
links\cite{Schaefer:2010hu}; we expect this behavior also in the theory
with fermions. Typical domain decompositions lead to active link ratios
between $0.37$ and $0.5$ and therefore a factor between 2 and 3 increased
autocorrelation times.

For some lattices, we therefore used a Hybrid Monte Carlo
algorithm\cite{Duane:1987de}
with a mass preconditioned fermion determinant\cite{algo:GHMC,algo:GHMC3}.
Our implementation\cite{Marinkovic:2010eg},  MP-HMC from here on, is based on the DD-HMC
package and in particular takes over the deflated solver because of its
efficient light quark inversions. This algorithm was employed for ensembles
A5, N6 and O7 given in Table~\ref{tab:ens}; all other ensembles were
generated with the DD-HMC. Appendix~\ref{a:1} gives details about these 
algorithms and the values of the parameters used for the gauge field
generation.

\subsection{Autocorrelations\label{sec:ac}}
In Monte Carlo data the effect of the autocorrelations has to be
accounted for in the error analysis. 
For all observables $F=F(a_1,\dots,a_n)$, functions of expectation
values $a_i=\langle A_i \rangle$ of primary observables $A_i$, 
we therefore compute an estimator of the autocorrelation function
\be
\Gamma_F(t)=\sum_{i,j} f_i(\vec a) f_j(\vec a) \langle (A_i(t)-a_i)
   (A_j(0)-a_j) \rangle
\ee
where $f_i=\partial_i F(\vec a)$ following the procedures detailed in
Ref.~\cite{Wolff:2003sm}. The argument $t$ indicates the Monte Carlo
time. The integrated autocorrelation time is then
\be
\tauint(F)=\frac{1}{2}+\sum_{t=1}^\infty
\frac{\Gamma_F(t)}{\Gamma_F(0)}\ .
\label{eq:tauint}
\ee
 which then enters the statistical error  of the observable $\sigma_F$ from $N$ 
measurements 
\be
\sigma_F^2 = 2\frac{\tauint(F)}{N}  \Gamma_F(0)\ . 
\ee

The sum in \eq{eq:tauint} is normally truncated at a ``window''
$W$\cite{Madras:1988de} which balances the statistical uncertainty due
to the limited sample size and the systematic error coming from
neglecting the tail for $t>W$.  The value of $W$ is determined from  the
measurement of $\rho_F(t)=\Gamma_F(t)/\Gamma_F(0)$ alone and for each
$F$ separately. Neglecting the tail above $W$ leads --- at least on
average --- to an underestimation of $\tauint$ and the statistical error
of the observable.  It is particularly problematic in the presence of
slow modes of the Monte Carlo transition matrix which only couple weakly
to the observable in question. To account for them we use the method
outlined in \Ref{Schaefer:2010hu}, estimating their time constants  from
observables to which the slow modes couple strongly. Using them, we can
then estimate the tails of the autocorrelation functions of the
observables we are interested in and arrive at a more conservative error
estimate.

Experience tells us that for small lattice spacing the topological
charge is particularly sensitive to slow
modes\cite{Schaefer:2009xx,Schaefer:2010hu}, for which we use the field 
theoretical definition after smoothing the field by the Wilson
flow integrated up to $t_0$ as defined in \Ref{Luscher:2010iy}.
Actually, only the square of the  charge needs to be considered, because
we are only interested in parity even observables. 
Unfortunately we are not in the position to accurately determine its
autocorrelation time for most of our ensembles. We therefore combine the
scaling laws found in pure gauge theory\cite{Schaefer:2010hu} with the
measurement for our high statistics ensembles E5 and arrive at the
estimate 
\be
\tau_\mathrm{exp}(\beta)=200\, {c_\tau\, e^{7(\beta-5.5)}\over R_{\rm active}} \, ,
\ee
in units of molecular dynamics time
with $c_\tau=2$ for trajectories of length $\tau=0.5$ and $c_\tau=1$
for $\tau=2$ and 4. The values of  $R_\mathrm{active}$ can be found in
\tab{tab:ddhmc} for the DD-HMC algorithm and is equal to one for the
MP-HMC.

\begin{figure}[hbt!]
\begin{center}
\includegraphics{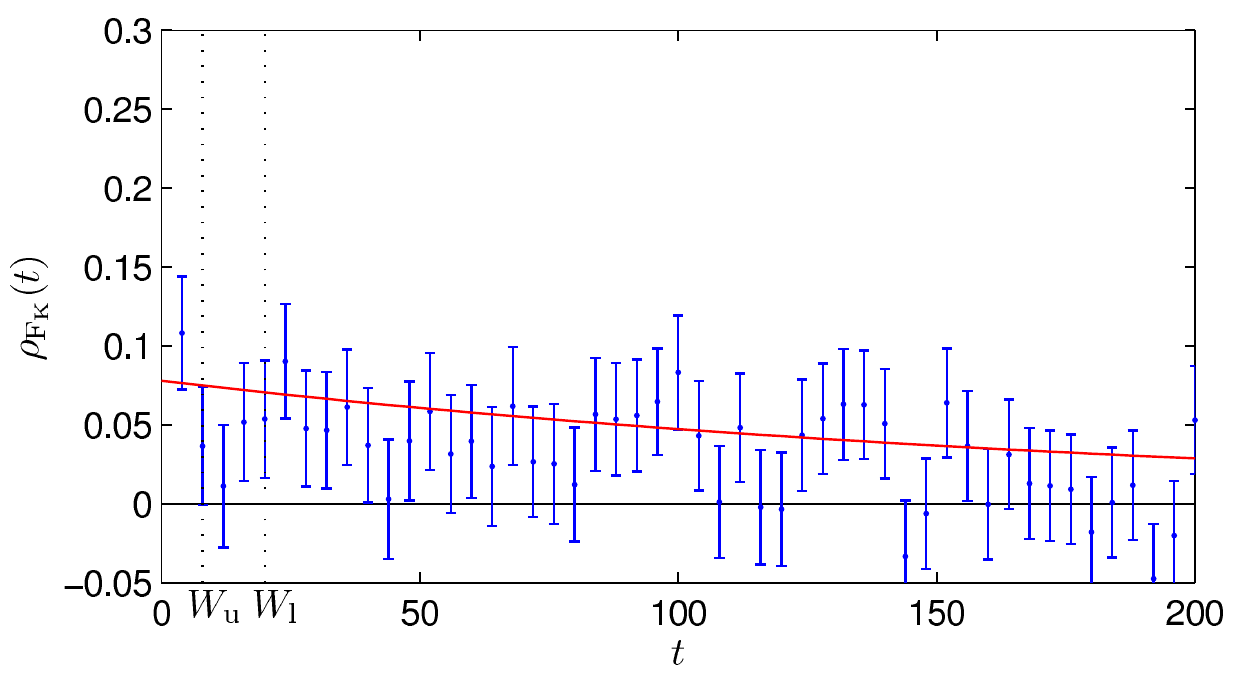}
\end{center}
  \caption{\footnotesize
    Autocorrelation function of $\fklat$ for the O7 lattice. The
       line gives our estimate for its tail. The standard method of
       \Ref{Wolff:2003sm} gives a window $W=W_l$ and
       $\tau_\mathrm{int}=0.7$, compared to
       $\tau_\mathrm{int}=4$ including the tail contribution which we
       add from $W=W_u$, or  more than a factor two in the error of the observable.
  }\label{f:tail}
\end{figure}

An example of the procedure is given in \fig{f:tail}, showing the 
autocorrelation function of the kaon decay constant $\fklat$ on the 
O7 ensemble. Using the standard
procedure\cite{Wolff:2003sm,Madras:1988de}, the sum in 
\eq{eq:tauint} is truncated at the window $W_l$ from which we
would get $\tau_\mathrm{int}=0.7$. When the contribution of the  tail is
included, the improved estimate gives $\tau_\mathrm{int}=4$, which 
translates to a more than doubled error estimate.

\section{Observables\label{s:3}}
The kaon decay constant necessarily requires the addition of a quenched
strange quark to the $\nf=2$ theory. We denote the hopping parameter of
this third flavor as $\kappa_3$ and have
$\kappa_1=\kappa_2=\kappa_\mathrm{sea}$ for the two sea quarks. For the
purpose of a definition of the strange quark mass from the PCAC
relation, we do in fact add a fourth quenched flavor with
$\kappa_4=\kappa_3$.

The computation of our pseudoscalar observables
is based on two-point functions of the
pseudoscalar density and the time component of the 
axialvector current. At a fixed 
$\kappa_\mathrm{sea}$ they are constructed 
from two valence quarks $r$
and $s$
\begin{align}
f_\mathrm{PP}^{rs}(x_0)&=-a^3\sum_{\vec{x}} \langle P^{rs}(x) P^{sr}(0) \rangle\\
f_\mathrm{AP}^{rs}(x_0)&=-a^3\sum_{\vec{x}} \langle A_0^{rs}(x) P^{sr}(0) \rangle
\label{eq:2pt}
\end{align}
with $P^{rs}=\psibar_r \gamma_5 \psi_s$ and $A_0^{rs}=\psibar_r
\gamma_0\gamma_5 \psi_s$. This notation and the analysis that follows is
similar to the one presented in Ref.~\cite{cern:II}. 
Using the PCAC relation, average quark masses
of flavors $r$ and $s$ can then be defined as\footnote{This definition
differs by a factor of two from Ref.~\cite{cern:II}.}  
\be
m_{rs}(x_0)=\frac{\frac{1}{2}(\partial_0 +\partial_0^*) 
  f_\mathrm{AP}(x_0)+ \ca a
  \partial_0^* \partial_0 f_\mathrm{PP}(x_0)}{2 f_\mathrm{PP}(x_0)}\ .
\label{eq:m}
\ee
In this formula, $\partial_0$ and $\partial_0^*$ denote the forward and
backward difference operators in time direction. The
improvement coefficient $\ca$ has been determined
non-perturbatively\cite{impr:ca_nf2}.

For sufficiently large $x_0$ the mass $m_{rs}(x_0)$ will have a plateau over
which we can average. From its value $m_{rs}$  the renormalized quark
mass $\mr^{rs}$ is obtained\cite{impr:nondeg}
\be
  \mr^{rs}=\frac{\za (1+\babar a \msea +  \batil a m_{rs})}
              {\zp (1+\bpbar a \msea +  \bptil a m_{rs})} m_{rs}\,,
\label{eq:mr}
\ee
with $\msea=m_{12}$.
We will use one-loop perturbation theory for the improvement coefficients
$\babar$, $\bpbar$, $\batil$ and  $\bptil$, noting that they multiply very small
terms. At this order in perturbation theory $\babar=\bpbar=0$, while
$\batil=1+0.06167\,g_0^2$ and $\bptil=1+0.06261\,g_0^2$ computed
from the perturbative 
coefficients of \cite{impr:pap5}. 
An update of the  non-perturbative determination of
$\za$~\cite{DellaMorte:2008xb} and $\zp$ \cite{mbar:nf2} 
is given in \app{a:ZA} and \app{a:ZP}, respectively.

The renormalization and improvement
of the PCAC quark masses is much simpler than the corresponding
expression in terms of the bare subtracted 
quark masses $m_{\mathrm{q},r}=m_{0,r}-m_\mathrm{cr}$, where terms proportional to 
$\msea$ are present already at the leading order in $a$ \cite{impr:nondeg}. 
In our analysis we therefore only use
the renormalized PCAC relation \eq{eq:mr}.
The alternative definition of renormalized quark masses
as well as the determination of additive renormalization
$m_\mathrm{cr}$ and the multiplicative renormalization factor
$Z_m$ is discussed in \app{a:mcrit}.

\subsection{Computation of the two-point functions}
We compute the two-point functions \eq{eq:2pt} using $U(1)$ noise
sources $\eta_t(x)=\delta_{t,x_0}\exp(i\phi(\vec x))$
 located on randomly chosen time slices
$t$\cite{Sommer:1994gg,Foster:1998vw}. 
Solving the Dirac equation once for each noise vector 
$\zeta_t^r=Q^{-1}(m_{0,r}) \eta_t = a^{-1}(D+m_{0,r})^{-1}\gamma_5 \eta_t$ 
is sufficient
to get an estimator for the two-point functions projected to zero
momentum
\bes
a^3 f_\mathrm{PP}^{rs}(x_0)&=&\sum_{\vec x}\langle 
   [\zeta_t^r(x_0+t,\vec x)]^\dagger \zeta_t^s(x_0+t,\vec x)\rangle\,,
\\
a^3 f_\mathrm{AP}^{rs}(x_0)&=&\sum_{\vec x} \langle
   [\zeta_t^r(x_0+t,\vec x)]^\dagger \gamma_0 \zeta_t^s(x_0+t,\vec x) \rangle
\ees
where the average is over noise sources and gauge configurations.
For our lattices, we use 10 noise sources per configuration,
balancing the numerical cost and the accuracy which we wanted to
reach on the given ensembles.

\subsection{Analysis of the data\label{s:dana}}
The following presentation applies to any flavor combination ``$rs$''
and we drop this sub/superscript for the sake of brevity.
The mass of the pseudoscalar meson $m_\mathrm{PS}$ and its decay constant
$f_\mathrm{PS}$ can be extracted from $f_\mathrm{PP}$ and the PCAC mass. 
For infinite time extent $T$,
the spectral decomposition gives an expansion in terms of functions
which decay exponentially for large time separations
\be
f_\mathrm{PP}(x_0)= \sum_{i=1}^\infty c_i e^{-E_i x_0} 
\label{eq:spec}
\ee
with $E_1=m_\mathrm{PS}$ the energy of the ground state and $E_2<E_3<\dots$ the
excited state contribution.
For large time separations, we can then
extract the decay constant from the leading coefficient
\bes
  f_\mathrm{PS} &=& \za (1+\babar a \msea +a  \batil m_{rs})
\,f_\mathrm{PS}^\mathrm{bare}\,,
  \label{e:fpsr}
  \\
  f_\mathrm{PS}^\mathrm{bare}&=& 
  2 \sqrt{2c_1} m_{rs} m_\mathrm{PS}^{-3/2} \ .
\ees

\begin{figure}
\begin{center}
\includegraphics{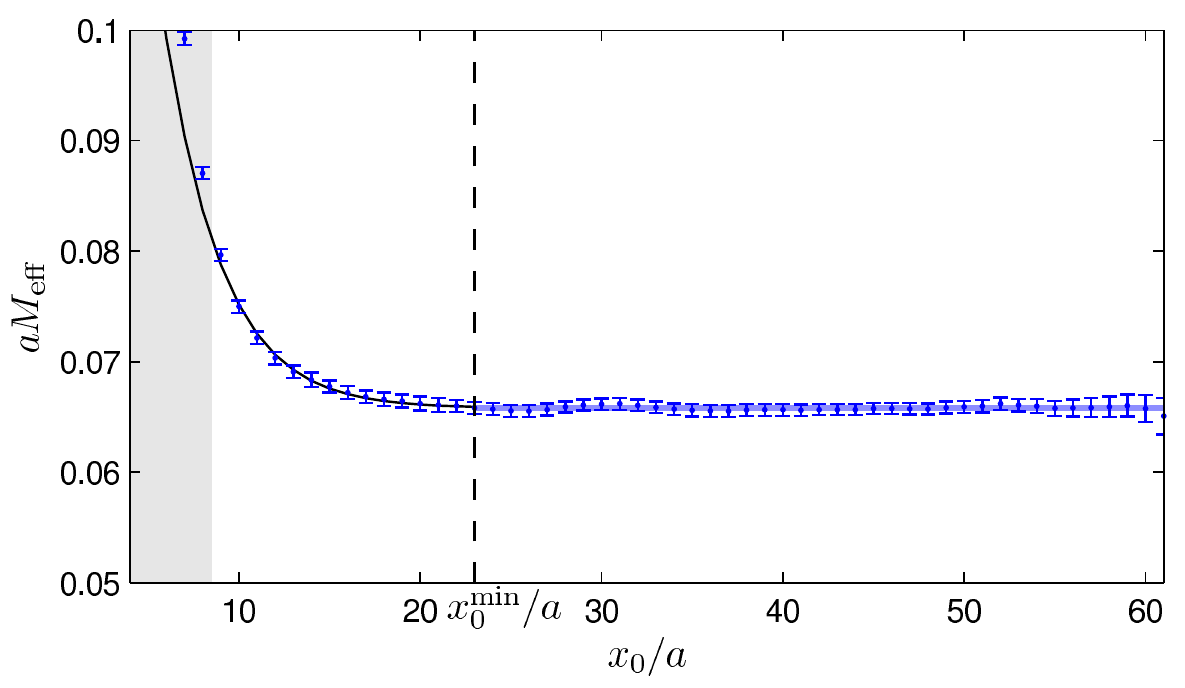}
\end{center}
\caption{\footnotesize  The effective pion mass given by 
   $\cosh(M_{\rm eff}(t-T/2))/\cosh(M_{\rm eff}(t+1-T/2))=f_\mathrm{PP}(t)/f_\mathrm{PP}(t+1)$
      for the O7 lattice. A two-state fit to data outside the shaded area determines 
      $x_0^\mathrm{min}$.  The result of the final one-state fit is given by 
      the error band. \label{f:ecitedstateO7}}
\end{figure}

In the analysis with finite time extent $T$ and time separation $x_0$
we have to deal with particles running backwards in time and excited
states.  Since our lattices are large and the statistical
precision of pseudoscalar correlators does not deteriorate
at large $x_0$, we use the following procedure
to fix the region $x_0 \in [x_0^\mathrm{min},T-x_0^\mathrm{min}]$,
in which we can neglect the excited state
contribution: we first perform a fit to the data using the first 
two terms in the expansion \eq{eq:spec}, now including the finite $T$
effects
\be
f_\mathrm{PP}(x_0)=c_1 \big[e^{-E_1x_0}+e^{-E_1(T-x_0)}\big]+c_2
\big[e^{-E_2x_0}+e^{-E_2(T-x_0)}\big]
\label{eq:spec2}
\ee
to a range where this function describes the data well, given the
accuracy of the data. We then determine $x_0^\mathrm{min}$:
it is the smallest value of $x_0$ where the statistical uncertainty on
the effective mass is four times larger than the contribution of the
excited state as given by the result of the fit using
\eq{eq:spec2}. In a second step, only the first term of \eq{eq:spec2} is
fitted to the data restricted to this region.
\Fig{f:ecitedstateO7} illustrates the procedure on our largest lattice.
Formally at large $x_0$ and small sea quark mass the leading correction to
\eq{eq:spec} comes from states which additionally to the 
ground state have two pions, $E_2\approx E_1+2m_\pi$.
 However, for small quark masses
and large $L$, the coefficient $c_2$ can be computed in chiral perturbation 
theory. It turns out to be very small \cite{threepi:obar} (at least for our
large volumes) and such a 
contribution is invisible within our precision. In this sense the value
of $E_2$
determined by the fit may actually be a higher state, which is one reason for us
to use this excited state fit only in order to determine a safe $x_0^\mathrm{min}$.

The error on
the decay constant and the meson mass $m_\mathrm{PS}=E_1$ are computed including
the autocorrelations as described in Sec.~\ref{sec:ac}.
In appendix~\ref{app:a2} we give the partially quenched meson masses,
decay constants and quark masses as well as the values of $x_0^\mathrm{min}$,
which we use in our analysis.

\section{Scale parameter $r_0$  \label{s:r0}}

The analysis strategy for the scale $R_0(\beta,\kappa)=r_0/a$
\cite{pot:r0} is based on \cite{Donnellan:2010mx} and we refer to this work
for more detailed explanations and notation. The procedure consists of
measuring on-axis Wilson loops $W(r,t)$ on smeared gauge configurations,
extracting the static potential $V(r)$ and finally solving the equation
\be
\left. r^2\,F(r)\right|_{r=r_0} = 1.65 \,, \label{r0}
\ee
where $F(r)=V^\prime(r)$ is the static force. For the latter we use an
improved definition which eliminates cut-off effects at tree level.
Wilson loops are measured on gauge configurations after all links are
replaced by HYP smeared \cite{HYP} links. We take the
HYP2 parameter choice \cite{DellaMorte:2005yc}
$\alpha_1=1.0$, $\alpha_2=1.0$ and $\alpha_3=0.5$ and
do one HYP-smearing level.
The Wilson loops can be exactly represented as an observable in a theory
including static quarks and this first smearing step corresponds to the choice
of the static quark action (as far as the time-like links are concerned). On
the HYP2-smeared gauge link configurations we measure a correlation matrix of
Wilson loops $C_{lm}(t)$ for fixed $r$ by smearing the space-like links using
numbers $n_l$ and $n_m$ of spatial HYP smearing iterations.
Spatial HYP smearing means that
only staples restricted to spatial directions are used and we therefore need
only two parameters, which we set to $\alpha_2=0.6$ and $\alpha_3=0.3$. 
This second smearing step corresponds in the Hamiltonian formalism to the
construction of a variational basis of operators $\hat{O}_l$ that create a state
consisting of a static quark and anti-quark pair. We use a basis of
the operators, labelled by $l=1,2,3$. The numbers of
smearing iterations at each level $n_l$ are listed in \tab{tab:HYPlevels}. 
They are chosen such that the physical
extensions of the operators are approximately constant as the lattice
spacing changes.
Finally we use the generalized eigenvalue method to extract the static
potential from the correlation matrix $C_{lm}(t)$, the details of this can be
found in \cite{Donnellan:2010mx}.

\begin{table}[t]
\small
\begin{center}
\begin{tabular}{@{\extracolsep{0.3cm}}cccc}
\toprule
$\beta$ & $n_1$ & $n_2$ & $n_3$ \\
\midrule
5.2 & 6  &  9 & 15 \\
5.3 & 8  & 12 & 20 \\
5.5 & 16 & 24 & 40\\
\bottomrule
\end{tabular}
\end{center}
\caption{\footnotesize The number of smearing levels $n_l$ used to construct the Operators
$\hat{O}_l$, $l=1,2,3$ for measuring the Wilson loop correlation
matrix.  \label{tab:HYPlevels}
}
\end{table}

\begin{table}[t]
\small
\begin{center}
\begin{tabular}{@{\extracolsep{0.3cm}}ccllll}
\toprule
$\beta$ & $S_2\equiv0$, $c=1.1$ & $S_2\equiv0$, $c=1.4$ & $c=1.4$ & $c=1.8$ \\
\midrule
$5.2$ & $6.145(59)$ & $6.119(45)$ & $6.22(16)$ & $6.160(79)$ \\
$5.3$ & $7.259(66)$ & $7.226(47)$ & $7.33(17)$ & $7.274(83)$ \\
$5.5$ & $10.00(11)$ & $9.953(90)$ & $10.10(26)$& $10.04(13)$ \\
\bottomrule
\end{tabular}
\end{center}
\caption{\footnotesize Chiral extrapolations of $R_0$ from global fits of the
form \eq{globalfitr0}. The cut $c$ indicates the upper bound in $x$; the
fits with  $S_2 \equiv 0$ do not include a quadratic term in $x$.
The final results are taken from the first set.
\label{tab:r0chiral}}
\end{table}

The solution of \eq{r0} is found by interpolation of the force $F$,
using a 2-point interpolation $F(r)=f_0+f_2/r^2$. In order to control the
systematic error we compare the result with a 3-point
interpolation adding a $f_4/r^4$ term. We find the systematic error to be
negligible. The error analysis takes into
account the coupling to the slow modes as explained in \sect{sec:ac} and we
neglect the systematic error due to excited state contributions to the
potential, because  we ensure that it is much smaller than
the statistical one. The values of $R_0$ for each ensemble are listed in 
\tab{tab:ens}.
\begin{figure}[t]
  \begin{center}
   \includegraphics{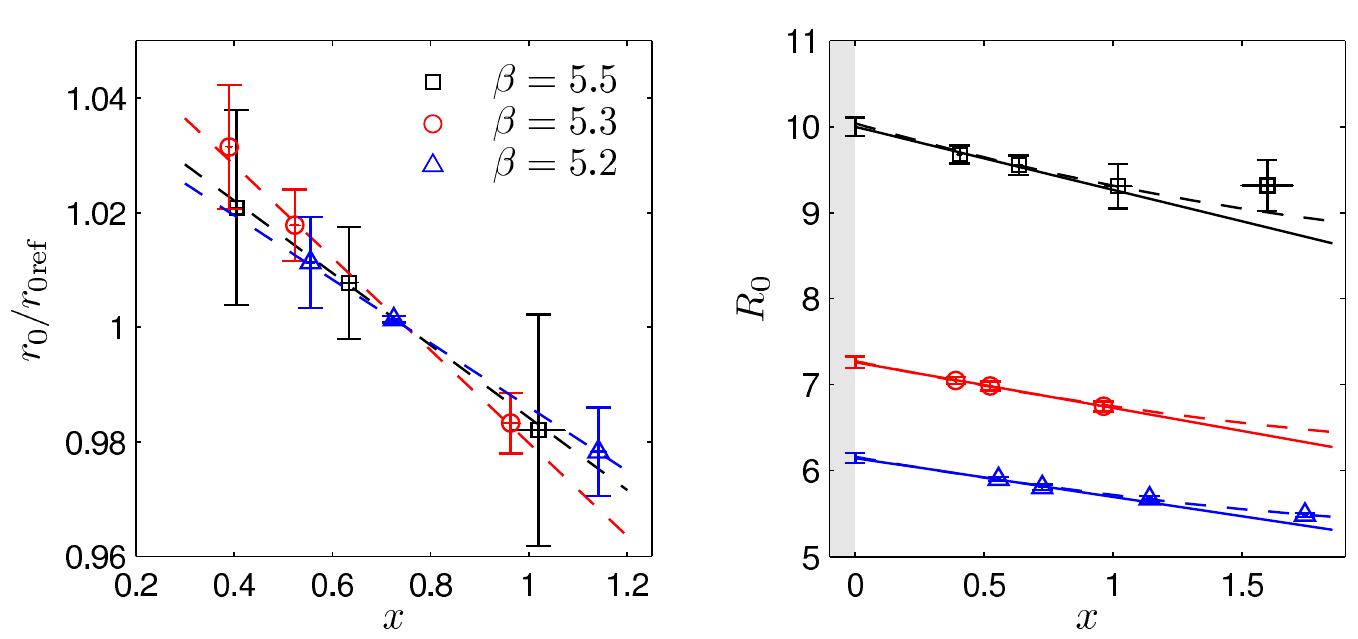}
  \end{center}
  \caption{\footnotesize The left plot shows the ratio $r_0/\r0ref$ as a
    function of $x = (r_0 m_{\rm PS})^2$, where
    $\r0ref$ is defined at the reference point \eq{xref},
    and the cut $x\le1.4$ is applied.
    The right plot shows our data of $R_0=r_0/a$ with the chirally
    extrapolated values using a linear function in $x$ (solid lines) applying
    the cut $x\le1.1$. For comparison a quadratic fit with $c=1.8$ is also
    shown (dashed lines).}
  \label{fig:r0}
\end{figure}

In order to define the lattice spacing in a mass independent way we need to
perform a chiral extrapolation of $R_0$ at fixed $\beta$. The chiral fits
are done in the variable
\be
x = (r_0 m_{\rm PS})^2 \,.
\ee
Before discussing the chiral fits we study the cut-off effects in the mass
dependence of $r_0$. For this purpose we
define a reference value $\r0ref$, which corresponds to the value of
$r_0$ at the pseudoscalar mass
\be
x_{\rm ref} \equiv (r_0 m_{\rm PS})^2|_{\rm ref} = 0.75 \,. \label{xref}
\ee
The reference point \eq{xref} corresponds to a pseudoscalar mass of
$340$\,MeV.
In the left plot of \fig{fig:r0}, we plot $r_0/\r0ref$ versus $x$ at
our three different $\beta$ values.
The value of $\r0ref$ is obtained by linear interpolation and the
error analysis of $r_0/\r0ref$ takes into account the correlations
between the data. We only consider data with $x\le1.4$.
Using these data
we determine the first coefficient in the Taylor expansion of
$r_0/\r0ref(x)$ around $x_{\rm ref}$ independently for each value of
$\beta$
\be
\frac{r_0}{\r0ref}(x) = 1 + s(a/\r0ref) \cdot (x-x_{\rm ref})
\,. \label{slopes}
\ee
We do not find significant cut-off effects in the slope $s(a/\r0ref)$
\cite{Knechtli:2011pz} and in the continuum limit we obtain
the value $s(0)=-0.067(10)$ by fitting to a constant.

Motivated by the results in the left plot of \fig{fig:r0}, we perform a global
fit for all $\beta$ values simultaneously of the form
\be
R_0(\beta,x) = \R0c(\beta)\left(1 + S_1\,x + S_2\,x^2\right) \label{globalfitr0}
\ee
and with cuts 
\be
x \le c
\ee
on the pseudoscalar mass.
The cuts $c=1.1\,,1.8$ correspond to 
$m_{\rm PS}=410\,\MeV$, $530\,\MeV$.
The fit takes into account both errors on $R_0$
and on $x$. The chirally extrapolated values
$\R0c(\beta)$ are listed in \tab{tab:r0chiral} for various
possibilities of the global fit.
Columns two and three are linear fits while columns four and five are quadratic
fits. We quote as our final numbers for $\R0c(\beta)$ the results in the second
column of \tab{tab:r0chiral} from the linear fit with $c=1.1$ (which has
$\chi^2/{\rm dof}=0.07$ and $S_1=-0.073(14)$):
\begin{align}
\R0c(5.2)&=6.145(59)\,,&
\R0c(5.3)&=7.259(66)\,,&
\R0c(5.5)&=10.00(11)\,. \label{R0c}
\end{align}

The data for $R_0(\beta,x)$ and the linear fit with $c=1.1$ (solid lines)
are shown in the right plot of \fig{fig:r0}.
The numbers in \eq{R0c} cover all the fit results of
\tab{tab:r0chiral} within errors. 
In particular they are perfectly consistent with the results in the fifth
column of \tab{tab:r0chiral} from a quadratic fit applying the cut $c=1.8$ 
(which has $\chi^2/{\rm dof}=0.28$, $S_1=-0.085(25)$ and
$S_2=0.013(11)$). This quadratic fit is represented by dashed lines in the
right plot of \fig{fig:r0}. The quadratic term is not 
significant even with $c=1.8$.

The discrepancy with the determination of $r_0$ by QCDSF
\cite{Brommel:2006ww} was discussed in \cite{Leder:2010kz}. Meanwhile QCDSF
updated their values,\footnote{G. Bali, private communication.}
which now agree with our determination where they can be compared.

\section{Chiral extrapolation of $\fk$ and strange quark mass\label{s:5}}

In this section we describe our central determination of
the scale. For a number of reasons, it is based on the kaon 
decay constant $\fk$. First, $\fk$ is experimentally accessible
once the CKM-matrix element $V_\mathrm{us}$ is considered
known, which is a good assumption within the envisaged 
precision. Second, chiral perturbation theory (ChPT)
provides a theory for the quark mass dependence
of $\fk$ at small masses of the light quarks, i.e.,
our sea quarks. Third, as already mentioned, we remain 
within the pseudoscalar sector of the theory, where 
ground state properties can be determined without doubt.

\subsection{Our strategies}
The main difficulty and source of a systematic error
is the extrapolation to the proper quark masses,
the ``physical point''. Once we decide to set the
scale through $\fk$, this point is naturally defined
by 
\bes
    R_\mathrm{K} =  R_\mathrm{K}^\mathrm{phys}\,, \label{e:rkpiphys}
    \quad
    R_\mathrm{\pi} =  R_\mathrm{\pi}^\mathrm{phys}\,, 
\ees
where 
\bes  
   R_\mathrm{K} = {\mk^2 \over \fk^2}\,, \quad
   R_\mathrm{\pi} = {\mpi^2 \over \fk^2} \,,
\ees
and $R_\mathrm{K}^\mathrm{phys},\; R_\mathrm{\pi}^\mathrm{phys}$
are the values of these ratios in Nature. In an attempt to
minimize uncertainties, we take the physical masses and decay
constants to be the ones in the isospin symmetric limit with QED 
effects removed as discussed in \cite{Flag1}. We use
\begin{align}
\mpiphys&=134.8\,\MeV\,,&
\mkphys&=494.2\,\MeV\,,&
\fkphys&=155\,\MeV\,.
\end{align}

The two conditions \eq{e:rkpiphys} define a point in the plane 
spanned by
$\mlight$, $\mstrange$, or equivalently $\kappa_1$, $\kappa_3$.
We are presently not able to simulate at or very close to this physical 
point, especially not for the smallest lattice spacing, where huge lattices
would be needed to keep finite size effects under control. The 
physical point has to be approached from unphysically large values of 
$\mlight$. Along which trajectory in the plane of bare parameters
one approaches the physical point 
is in principle arbitrary. However, one would like the 
quantity that is to be computed --- here $\fk$ --- to depend very
little on the distance to the physical point, allowing for an
easy extrapolation. Secondly, one would like a theoretically motivated
extrapolation formula. In ChPT both $\mk$ and $\fk$ depend on the sum of the
quark masses at leading order in the systematic expansion
in small quark masses. Keeping 
\bes
   R_\mathrm{K}(\kappa_1,\kappa_3) =   R_\mathrm{K}^\mathrm{phys} 
   \label{e:Rkcond}
\ees
thus defines a trajectory where $\fk$ varies little in ChPT. 
We will discuss this quantitatively below. An additional advantage
is that all along this trajectory we
have $\mk\approx\mk^\mathrm{phys}$, while for a more conventional trajectory, 
where $\mstrange$ is kept constant, the mass $\mk$ is significantly heavier 
than $\mk^\mathrm{phys}$. 
Since the ChPT expansion is written in terms 
of $\mk^2$ and $\mpi^2$, having $\mk$ no larger then $\mk^\mathrm{phys}$ 
increases our chance of being inside the expansion's
domain of applicability.
To our knowledge this strategy has not been used so far, somewhat surprisingly.

As an alternative we use a second strategy, where we keep 
$\mstrange$ constant, however instead of using the expansion in 
$\mk^2$ and $\mpi^2$, we use SU(2) ChPT 
for kaon observables \cite{HKCHPT:roessl,HKCHPT:chris},
where only $\mpi^2$ is considered a small parameter (in units of the chiral
scale $4\pi f$). This strategy provides  a suitable definition for 
future investigations of mesons and  baryons with strange quark content.

\begin{figure}[tb!]
\begin{center}
\includegraphics{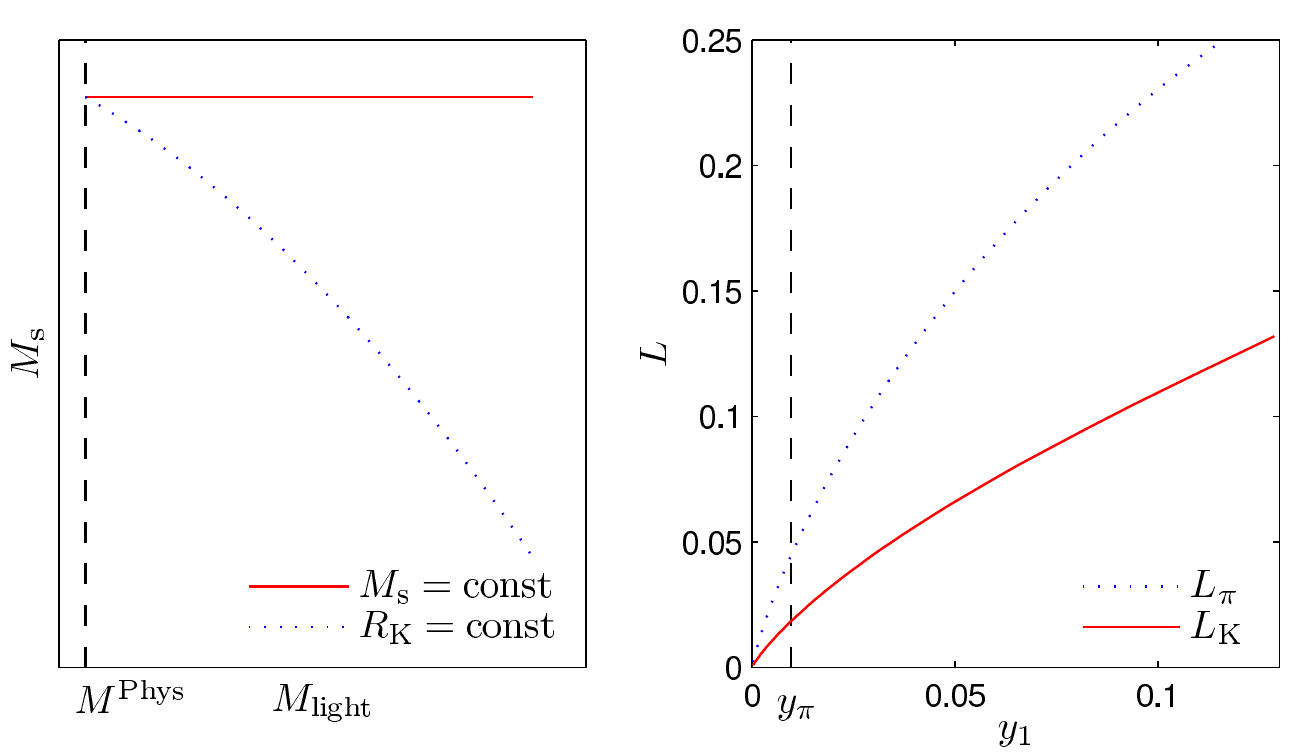}
\end{center}
  \caption{\footnotesize
     Left:
        Trajectories to approach the physical point in the
        plane of light and strange quark mass $\Mlight,\,\Mstrange$.
        The dotted line corresponds to strategy 1, i.e., $R_\mathrm{K} =
        R_\mathrm{K}^\mathrm{phys}$, whereas strategy 2 (full line) holds 
        $M_\strange$ fixed.
     Right: The two functions $L_\mathrm{K}$ and  $L_\pi$, \eqs{e:fkstrat1c} and
           (\ref{e:fpi}) respectively, in the interval 
	$y_1\in[0,\yk]$.
}\label{f:strategies}
\end{figure}

The trajectories belonging to the two strategies, whose details are the
subject of the following sections,  are schematically
shown in \fig{f:strategies} on the left.

Since the following section deals with the scale setting, we want to
distinguish between the kaon decay constant in physical units $\fk$ and
in lattice units $\fklat$
\be
\fklat\equiv a\fk\,.
\ee

\subsection{Strategy 1 and partially quenched SU(3) ChPT}
As discussed above, the trajectory
$ \kappa_3=h(\kappa_1)$ 
 is defined by 
\bes
   R_\mathrm{K}(\kappa_1,h(\kappa_1)) =  R_\mathrm{K}^\mathrm{phys}\,,
   \label{e:defhk3}
\ees
where $R_\mathrm{K}$ is considered a function
$R_\mathrm{K}(\kappa_1,\kappa_3)$\,.
In practice we take a fixed 
sea quark hopping parameter $\kappa_1$ and a few values
of $\kappa_3$ for which $R_\mathrm{K}(\kappa_1,\kappa_3)$
is close to $R_\mathrm{K}^\mathrm{phys}$ and then interpolate in
$\kappa_3$ to find $h(\kappa_1)$. Also $\fk(\kappa_1,\kappa_3)$
is interpolated to $\kappa_3=h(\kappa_1)$. We give more details 
on the various interpolations in $\kappa_3$ in \sect{s:interpol}.

It remains to extrapolate $\fk(\kappa_1,h(\kappa_1))$ in 
$\kappa_1$ to the physical point \eq{e:rkpiphys}. 
We use the 
partially quenched chiral perturbation theory results by
Sharpe \cite{PQChPT:Steve}, implement our condition
\eq{e:Rkcond}, which expresses $\mk$ in terms of $\mpi$,
and find 
\begin{align}
  \label{e:fkstrat1}
  \fk(\kappa_1,h(\kappa_1)) &= \fkphys\,\left[1 + \overline{L}_\mathrm{K}(y_1,\yk) 
  + (\alpha_4-\frac14)\,(y_1-\ypi) +\rmO(y^2) 
  \right],
  \\
  \overline{L}_\mathrm{K}(y_1,\yk) &=  L_\mathrm{K}(y_1,\yk) -
  L_\mathrm{K}(\ypi,\yk) \,, \label{e:fkstrat1b}
  \\
  L_\mathrm{K}(y_1,\yk) &= - \frac12 y_1 \log(y_1) - \frac18y_1\log(2\yk/y_1 - 1) \,.
  \label{e:fkstrat1c}
\end{align}
The variables $y_i$ are proportional to (averages of) quark masses up to 
quadratic terms
\begin{align}
 y_1 &= {\mpi^2(\kappa_1) \over 8\pi^2 \fk^2(\kappa_1,h(\kappa_1))}
 \,,\quad \nonumber \\
 \\\nonumber
 \ypi &= {\mpiphys^2 \over 8\pi^2 \fkphys^2}=0.00958\,, &
 \yk &= {\mkphys^2 \over 8\pi^2 \fkphys^2}=0.12875 \,.
\end{align}
Because of \eq{e:Rkcond}, we have 
$y_3\equiv{\mk^2(\kappa_1,h(\kappa_1)) / [8\pi^2 \fk^2(\kappa_1,h(\kappa_1))]}= 2\yk-y_1 +\rmO(y^2)$
and $y_3$ does not appear in \eq{e:fkstrat1}. 

At this order in the chiral expansion we can also replace
\bes
  y_1 \to \tilde y_1= {\mpi^2(\kappa_1) \over 8\pi^2 \fpi^2(\kappa_1)}\,,
\ees
with the corresponding replacement $\ypi \to \tilde y_\pi$,
which we will use as a check of the typical size
of $\rmO(y^2)$ effects. 

In the right panel of  \fig{f:strategies} we compare the
chiral log function $L_\mathrm{K}$ to the one describing the 
chiral behavior of $\fpi$,
\begin{align}
  \fpi(\kappa_1) &= \fpiphys\,\left[1 + \overline{L}_\pi(y_1) 
  + (\alpha_4+\frac12\alpha_5)\,(y_1-\ypi) +\rmO(y_1^2)
  \right] \,,
  \\
  \label{e:fpi}
  L_\pi(y_1) &= - y_1 \log(y_1) \,,\quad 
  \overline{L}_\pi(y_1) =  L_\pi(y_1) - L_\pi(\ypi) \,.
\end{align}
Our condition \eq{e:defhk3} leads to the specific combination of chiral
logarithms \eq{e:fkstrat1c}, which has very little curvature and is
overall much smaller than  $L_\pi$; the suppression of the light quark
mass dependence thus extends also to the NLO chiral logarithms. This
suggests that the chiral extrapolation is much easier than for $\fpi$
and was one of our reasons to select $\fk$ to set the scale.  Of course,
the counter terms $\alpha_i$ do not contribute in \fig{f:strategies} (right), but
as they are linear in $y_1$ introduce no curvature.

\subsection{Strategy 2 and SU(2) ChPT  \label{s:heavyK}}
Here we extrapolate in the light quark mass at fixed mass of the strange quark, namely 
we tune for each sea quark mass the strange quark's hopping parameter such that the 
PCAC mass $am_{34}$ has a prescribed value $\mu$, which is independent of 
$\kappa_1$.\footnote{In principle we should keep $\mr^{34}$ fixed,
but the ratio $\mr^{34}/m_{34}$ is independent of $\kappa_1$
since we set $\babar=\bpbar=0$, see \eq{eq:mr}.
}
This defines the function
$\kappa_3=s(\kappa_1,\mulat)$. In practice, we again 
interpolate the data for $m_{34}(\kappa_1,\kappa_3)$ in $\kappa_3$ and then solve 
\bes
   a m_{34}(\kappa_1,s(\kappa_1,\mu)) = \mulat \label{eq:kappas}
\ees
for $s$, with the left hand side represented by the interpolation formula.

To find the value of $\mulat$ corresponding to the physical point, 
we employ SU(2) ChPT 
\cite{HKCHPT:roessl,HKCHPT:chris}
to first extrapolate
$\mksqlat=(a\mk)^2$ and $\fklat$ (both interpolated in $\kappa_3$ 
to the point $\kappa_3=s(\kappa_1,\mulat)$) in $y_1$
to $y_1=\ypi$ at fixed value of $\mulat$,
\begin{align}
    \fklat(\kappa_1,s(\kappa_1,\mulat)) &= p(\mulat) \,\big[1 
  -\frac38 [y_1 \log(y_1) - \ypi \log(\ypi)] + \alpha_{\rm f}(\mulat)\,(y_1-\ypi) 
            + \rmO(y_1^2) \big]\,, \nonumber \\
    \label{e:hm}  \\\nonumber
    \mksqlat(\kappa_1,s(\kappa_1,\mulat)) &= q(\mulat) \,\big[1 
    + \alpha_{\rm m}(\mulat)\,(y_1-\ypi) + \rmO(y_1^2)\big]\,.
\end{align}
These expressions represent the asymptotic expansions for small $y_1$ at fixed 
$\mulat$ correct up to error terms of order $y_1^2$.\footnote{We note that 
SU$(2)$ ChPT for kaons does not take into account
kaon loops  as opposed to SU(3) ChPT. This corresponds to the production
of {\em two} kaons, i.e., states with an energy of around one 
GeV. From this point of view it is not really worse than ChPT 
in the pion sector where $|\rho\,\pi\rangle$ intermediate states are dropped. 
We thank Gilberto Colangelo for emphasizing this point. 
}

From \eq{e:hm}, $q(\mu)$ and $p(\mu)$ are computable
for arbitrary values of $\mu$.
The requirement that $m_\mathrm{K}^2/f_\mathrm{K}^2$ attains its physical 
value at the 
physical light quark mass then defines $\mustrange$,
\bes
  \label{e:mstrange}
  \frac{q(\mustrange)}{p(\mustrange)^2}
      = \frac{\mkphys^2}{\fkphys^2} \,. \label{eq:mustrange}
\ees
This equation is solved numerically for $\mustrange$ and the 
lattice spacing is then given by
\bes
  \label{e:aHM}
  a = {p(\mustrange) \over \fkphys}  \,.
\ees
As before, the constants 
$\alpha_{\rm f}(\mustrange), \alpha_{\rm m}(\mustrange)$ have
common values for all three $\beta$ in the fits to \eq{e:hm}. 

\subsection{Interpolations\label{s:interpol}}

In various places we need observables, such as
$R_\mathrm{K}(\kappa_1,\kappa_3)$ as a continuous function of
$\kappa_3$, not just for a few numerical values.  In all cases we chose
four different numerical values for $\kappa_3$, close to the required
one, namely those which have the smallest distance defined by
$d=|R_\mathrm{K}(\kappa_1,\kappa_3) - R_\mathrm{K}^\mathrm{phys}|$. We
then determine an interpolation polynomial $P_2(\kappa_3-\kappa_1) =
\sum_{n=0}^2 c_n\;(\kappa_3-\kappa_1)^n$ through a fit to the data with
weights $(d^2+\epsilon^2)^{-1}$. The regulating term $\epsilon$ is
chosen as
$
\epsilon=\frac{1}{100}\big
(\max_{\kappa_3}\{R_\mathrm{K}(\kappa_1,\kappa_3)\}-\min_{\kappa_3}\{R_\mathrm{K}(\kappa_1,\kappa_3)\}\big)\,.
$
We always
checked that uncertainties due to the specific interpolation are
negligible by considering natural variants.

\subsection{Cutoff effects\label{s:cutoff}}
\begin{figure}[tb!]
\vspace{0pt}
\centerline{\includegraphics{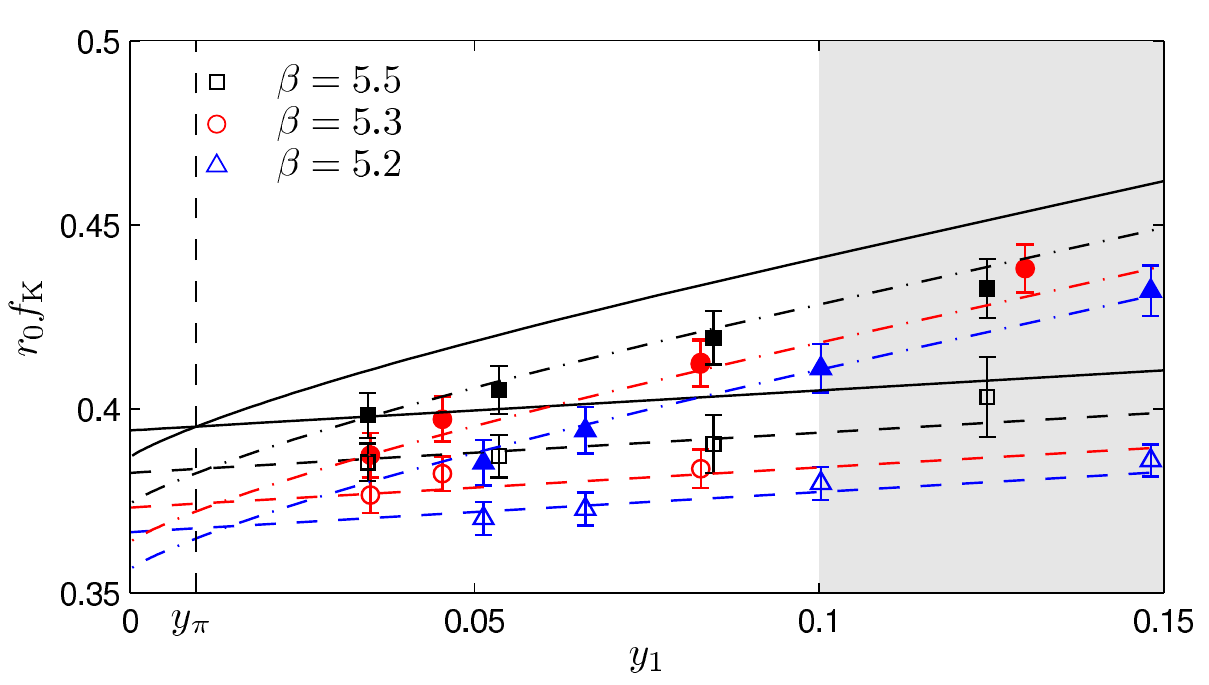}}
  \vspace*{-5mm}
  \caption{\footnotesize
     Chiral extrapolation of  $r_0\fk$. For the open symbols $r_0$ and
        $\fk$ are evaluated at finite quark mass, the dashed
        lines show \eq{e:r0fka} for the three values of $\beta$.
        The filled symbols use the extrapolated value $r_0=r_{0c}$, the 
         dashed dotted lines represent \eq{e:r0fkb}. In both cases, the
        corresponding solid line gives the continuum result. 
        \label{f:fkr0}}
\end{figure}

So far the discussion neglects lattice artifacts completely.
We will see that these are very small in our formulation.
Since both $y_1$ and $a^2$ are small, it is natural to keep 
terms of order $y_1$ (as done above) and those of order 
$a^2$ which are independent of $y_1$ but drop terms of
order $y_1^2,\,y_1 a^2,a^3$ etc. In this approximation we can use
\eq{e:fkstrat1} for the decay constants in lattice units
$\fklat$, and with a global, $a$-independent
low energy constant $\alpha_4$. The straight $a^2$ 
term does not contribute to \eq{e:fkstrat1}, since 
$\fkphys$ is used to set the scale. Such a term is, however,
present in general. An example is the combination  
$r_0 \fk$. It is shown in \fig{f:fkr0} with open symbols, 
where both $\fk$ and $r_0$ are evaluated at the finite quark mass.  
Since a ChPT expansion of $r_0$ does not exist, we use 
\bes
  \label{e:r0fka}
  R_0(\kappa_1)\,\fklat(\kappa_1,h(\kappa_1)) = 
  \left[r_0|_{y=\ypi}\,\fkphys\right]_\mathrm{cont}\,
  \left[1 + A\,(y_1-\ypi) + B \fklat^2\right] \,
\ees
to describe the data and to extrapolate to the physical point and continuum
limit $\left[r_0|_{y=\ypi}\,\fkphys\right]_\mathrm{cont}$. Taking into
account that we have $\fklat^2 < 4\cdot 10^{-3} $,
the lattice artifact parametrized by  $B$ is rather small, see \tab{t:chiralfits}.
In these fits we take the chirally extrapolated value $\fklat$ in the
last term of \eq{e:r0fka}. Replacing it with $\fklat$ at the given value
of $y_1$, we get an insignificant shift of $0.6\%$ in the continuum
value, which tests the smallness of $y_1a^2$ terms.

\begin{table}
\small
\begin{center}
\begin{tabular}{@{\extracolsep{0.2cm}}ccccccc}
\toprule
 $y_1^{\rm max}$ & $r_0|_{y=\ypi}\fkphys$ & $A$ &$ B$ & $r_\mathrm{0c}\fkphys$ & $\alpha_4$ & $B'$ \\
\midrule
$0.10$ &$ 0.3951(62)$&$ 0.28(22)$&$ -20(5)$& $ 0.3950(75)$ & $
0.52(11)$&$ -22(6)$\\
\bottomrule
\end{tabular}
\caption{\footnotesize Results for chiral fits in the range
 $y_1\leq y_1^{\rm max}$. \label{t:chiralfits}}
\end{center}
\end{table}

Alternatively, we use the already 
chirally extrapolated values  $\R0c$ from \sect{s:r0} and 
\begin{multline}
  \R0c\,\fklat(\kappa_1,h(\kappa_1)) \\
=  \left[r_\mathrm{0c}\,\fkphys\right]_\mathrm{cont}
  \left[1 + \overline{L}_\mathrm{K}(y_1,\yk) + (\alpha_4-\frac14)\,(y_1-\ypi) +
  B' \fklat^2\right] \,. 
  \label{e:r0fkb}
\end{multline}
The dashed dotted lines in \fig{f:fkr0} show the corresponding fit with  the parameters 
listed in \tab{t:chiralfits}.

The just discussed fits mainly illustrate that 
cut-off effects are small and provide a motivation that indeed terms
such as $a^2 y_1$ can be dropped. We then determine  
the values of the kaon decay constant in lattice units at the 
physical point $\fklatphys$ 
by a direct application of 
\bes
  \label{e:fkfit1}
  \fklat(\kappa_1,h(\kappa_1)) = \fklatphys\,\left[1 + \overline{L}_\mathrm{K}(y_1,\yk) 
  + (\alpha_4-\frac14)\,(y_1-\ypi) 
  \right]\,.
\ees
Here the data at the three different $\beta$ are combined
in a global fit with one common value of $\alpha_4$, but, of course, 
with different $\fklatphys$ for the three different $\beta$.
The lattice spacings
are then obtained from
\bes
  a = {\fklatphys \over \fkphys} \,.
\ees
Results are discussed in 
\sect{sec:6} together with our alternative strategy.

We finish the discussion of cut-off effects by mentioning 
an investigation of the PCAC relation. In a properly $\rmO(a)$
improved theory we have 
$\Delta=\frac12 (m^{12}_\mathrm{R} +m^{34}_\mathrm{R}) - m^{13}_\mathrm{R} =\rmO(a^2)$.
We have computed $\Delta$ for all our combinations of $\beta$,
   $\kappa_1$, $\kappa_3$ and found $a\Delta < 10^{-4}$
and $\Delta < 0.3\,\MeV$ in physical units.\footnote{The improvement term 
      proportional to $\batil-\bptil = (\ba-\bp)/Z$  was 
inserted non-perturbatively using $\ba-\bp$ and $Z$ from \cite{impr:babp_nf2}.
It is very small in practice.}
 This confirms again the
smallness of cut-off effects.

\subsection{\label{sec:6}Numerical results}

\begin{figure}[bt!]
\begin{center}
\includegraphics{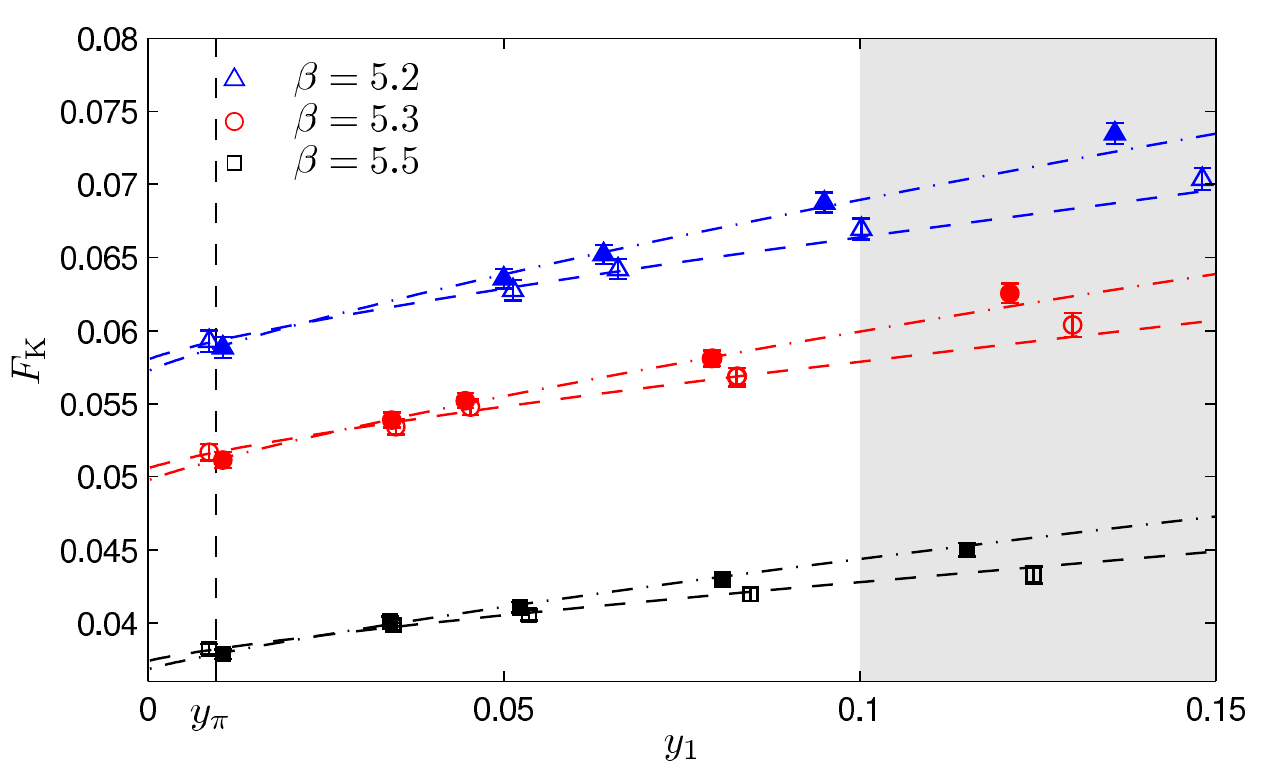}
\end{center}
  \caption{\footnotesize
    Chiral extrapolation of the kaon decay constant in lattice     
    units for all three $\beta$. Open symbols and dashed lines
    correspond to strategy 1, whereas filled symbols and    
    dash-dotted lines represent strategy 2.
  }\label{f:fkextrap1}
\end{figure}

We now apply the above formulae to a
determination of the kaon decay constant and the strange quark mass.
The renormalization of the 
decay constant, \eq{e:fpsr}, starts from $\za$ computed
in \cite{DellaMorte:2008xb} and improved in statistical precision
in \app{a:ZA}. For the mass-dependent $\rmO(a)$ improvement terms, 
which yield very small corrections, we use $\babar=0$ and
$\batil=1+0.06167g_0^2$
from one-loop perturbation theory \cite{impr:pap5}. 
The extrapolations of $\fklat$ through \eq{e:fkfit1}, strategy 1, and 
\eq{e:hm}, strategy 2, are shown in 
\fig{f:fkextrap1}. In the case of strategy 2, the parameter $\mu$ is fixed
to the strange quark mass $\mu=\mustrange$ through \eq{e:mstrange},
with the  values of $\kappa_\strange$  given in \Tab{tab:ens}.
The fits include data for $y_1<y_1^\mathrm{max}$ with $y_1^\mathrm{max}=0.1$. 
We observe that the two rather different chiral effective theory 
extrapolations yield results in close agreement at the physical point.
\Tab{t:chiralfits2} lists the fit parameters and also includes 
fits with $y_1 \to \tilde y_1$, simple linear fits in $y_1$ and
$\tilde y_1$ as well as results including data 
out to $y_1^\mathrm{max}=0.15$.

\begin{table}[t]
\begin{center}
\footnotesize
\begin{tabular}{@{\extracolsep{0.2cm}}ccccccccc}
\toprule
fit & $y_1^{\rm max}$ & $\beta$ & 
\multicolumn{3}{c}{fits with $y_1$} & 
\multicolumn{3}{c}{fits with $\tilde y_1$} \\
\cmidrule(lr){1-3}\cmidrule(lr){4-6}\cmidrule(lr){7-9}
\multicolumn{3}{c}{Strategy 1} &  $10^2\fklat$ & $\alpha_4$ & $10^2a$[fm]& $10^2\fklat$ & $\alpha_4$ & $10^2a$[fm] \\
\midrule
(\ref{e:fkfit1}) & 0.15 &
 5.2 &5.92(7) & 0.67(6) & 7.53(9) & 5.81(7) & 0.98(7) & 7.40(9)\\
 &  &
 5.3 &5.15(5) &  & 6.55(7) & 5.05(5) &  & 6.43(7)\\
 &  &
 5.5 &3.80(3) &  & 4.84(4) & 3.73(3) &  & 4.75(4)\\
     & 0.1 &
 5.2 &5.93(7) & 0.57(12) & 7.55(9) & 5.87(7) & 0.71(12) & 7.47(10)\\
 &  &
 5.3 &5.17(6) &  & 6.58(7) & 5.12(6) &  & 6.52(7)\\
 &  &
 5.5 &3.82(4) &  & 4.86(4) & 3.78(4) &  & 4.82(5)\\
linear & 0.1 &
 5.2 &5.99(7) & 1.24(12) & 7.62(9) & 5.93(7) & 1.26(12) & 7.55(9)\\
 &  &
 5.3 &5.21(6) &  & 6.64(7) & 5.17(6) &  & 6.58(7)\\
 &  &
 5.5 &3.85(3) &  & 4.91(4) & 3.82(4) &  & 4.86(5)\\
\midrule
\multicolumn{3}{c}{Strategy 2} & $10^2\fklat$ & $\alpha_{\rm f}$ & $10^2a$[fm]& $10^2\fklat$ & $\alpha_{\rm f}$ & $10^2a$[fm] \\
\midrule
(\ref{e:hm})   & 0.15 &
 5.2 &5.86(7) & 1.30(6) & 7.45(9) & 5.77(7) & 1.42(6) & 7.35(9)\\
 &  &
 5.3 &5.09(5) &  & 6.48(7) & 5.01(5) &  & 6.38(7)\\
 &  &
 5.5 &3.76(3) &  & 4.79(4) & 3.71(3) &  & 4.72(4)\\
     & 0.1 &
 5.2 &5.88(7) & 1.13(8) & 7.49(9) & 5.83(7) & 1.15(8) & 7.42(9)\\
 &  &
 5.3 &5.11(5) &  & 6.51(7) & 5.06(6) &  & 6.44(7)\\
 &  &
 5.5 &3.79(3) &  & 4.82(4) & 3.75(3) &  & 4.77(4)\\
linear & 0.1 &
 5.2 &5.97(7) & 1.78(8) & 7.61(9) & 5.92(7) & 1.72(8) & 7.54(9)\\
 &  &
 5.3 &5.19(5) &  & 6.61(7) & 5.15(5) &  & 6.55(7)\\
 &  &
 5.5 &3.84(3) &  & 4.89(4) & 3.81(3) &  & 4.85(4)\\
\bottomrule
\end{tabular}
\caption{\footnotesize Results for chiral fits in the range
 $y_1\leq y_1^{\rm max}$. 
\label{t:chiralfits2}}
\end{center}
\end{table}

A few observations are worth pointing out. 
Using $\tilde y_1$ as the chiral variable leads to a stronger
      dependence on the cut $y_1^\mathrm{max}$. Still,
      the extrapolation in terms of $\tilde y_1$ shows a tendency to
      converge to the one 
      in terms of $y_1$ when the cut is lowered, as it should be.
      Results of linear extrapolations are also very close. 

\begin{figure}[tb!]
\begin{center}
\includegraphics{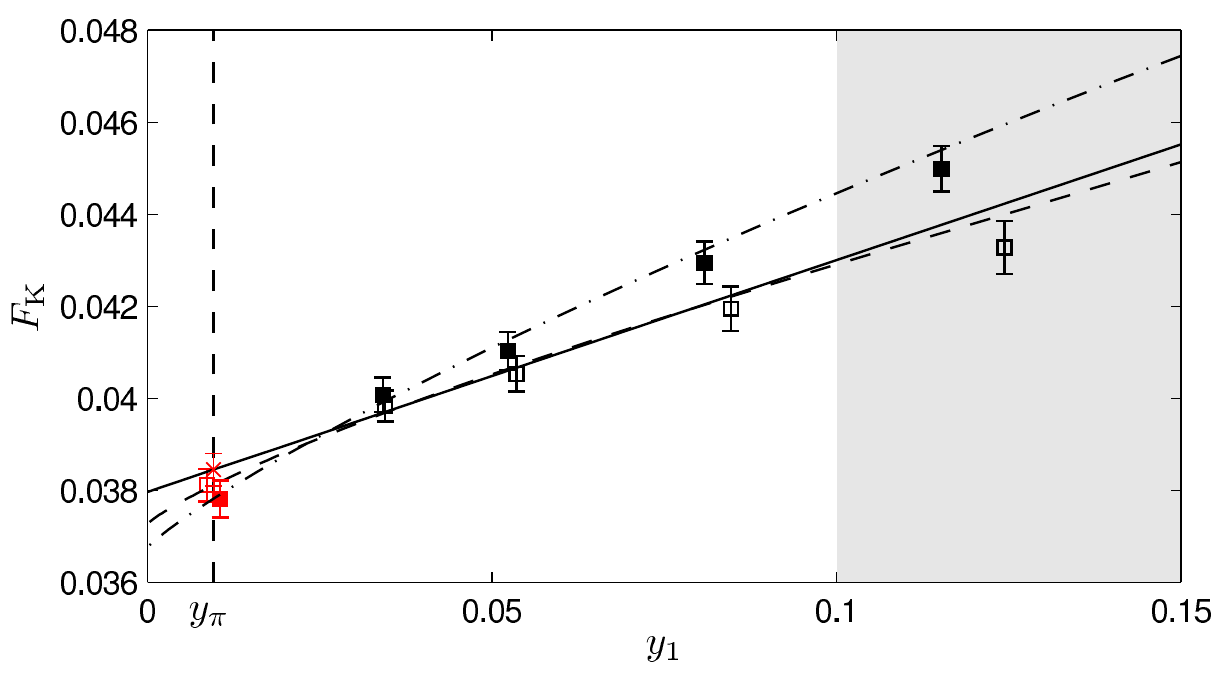}
\end{center}
  \caption{\footnotesize
        Detailed comparison of strategy 1, strategy 2 and a linear extrapolation
        at $\beta=5.5$. The symbols are the same as in \fig{f:fkextrap1}
        with the solid line representing the linear extrapolation of the
           open symbols. 
        Note that the slopes given by the low energy parameters $\alpha$ 
        are determined from the global fit to all three $\beta$.
        }\label{f:fkextrap2}
\end{figure}

We therefore take  our central values from  the results of strategy 1 with 
cut $y_1^\mathrm{max}=0.1$. As a systematic error we take into account
the deviations to the fit following strategy 2 and to a simple linear 
extrapolation; for our smallest lattice spacing these different extrapolations
are compared more closely in \fig{f:fkextrap2}. The final numbers 
for $\fklat$ at the physical point and the associated lattice spacings are
shown in \tab{t:chiralfits3}.

\begin{table}
\small
\begin{center}
\begin{tabular}{@{\extracolsep{0.2cm}}ccc}
\toprule
 $\beta$ & $\fklat$ & $a$[fm]\\
\midrule
 $5.2$ &$0.0593(7)(6)$ &  $0.0755(9)(7)$ \\
 $5.3$ &$0.0517(6)(6)$ &  $0.0658(7)(7)$ \\
 $5.5$ &$0.0382(4)(3)$ &  $0.0486(4)(5)$ \\
\bottomrule
\end{tabular}
\caption{\footnotesize
The kaon decay constant at the physical quark mass and the corresponding
lattice spacing. The  first error is statistical, the second systematic
error due to the  chiral extrapolation.
\label{t:chiralfits3}}
\end{center}
\end{table}

The fit parameters $\alpha_4$ and $\alpha_\mathrm{f}$ turn into low energy
constants of SU$(3)$ and SU$(2)$ chiral perturbation theory, respectively, when the cuts on $y$
are sufficiently small, such that the used NLO expressions are accurate. A look
in the tables shows that their dependence on the cut is anything  but
negligible. Clearly the values obtained with $y_1^{\rm max} = 0.1$ are our best
estimates. Differences to $y_1^{\rm max} = 0.15$ are an estimate of the
systematic error, which, however, deserves a more detailed study. We postpone
this to a later analysis when data at an even smaller quark mass will be
available for two values of the lattice spacing.

\section{Lambda parameter\label{s:6}}

The $\Lambda$ parameter, which normalizes the scales in the perturbative 
running of QCD quantities at high energies, has been determined in an 
earlier work by the ALPHA collaboration\cite{alpha:nf2} in terms of a
length scale
$L_1$.  It used the Schr\"odinger functional
which allows to compute the running over many
orders of magnitude in the scale $L$, and  $L_1$ is a typical
hadronic scale which allows to make contact to physical units.

In particular a renormalized
coupling $\gbar(L)$ can be defined  for which the renormalization
group invariant $\Lambda$ parameter is given by
\bes
L\Lambda= (b_0 \gbar^2(L))^{-b_1/(2 b_0^2)} e^{-1/(2 b_0 \gbar^2(L))}
\exp\left [-\int_0^{\gbar(L)} \!\mathrm{d}x \Big\{
   \frac{1}{\beta(x)}+\frac{1}{b_0x^3}-\frac{b_1}{b_0^2x}\Big
   \}\right].
\ees
Here $\beta(\gbar)=-L \partial \gbar(L)/\partial L$ with $b_0$ and 
$b_1$  the first two universal coefficients of its asymptotic expansion
$\beta(\gbar)=-\gbar^3(b_0+b_1 \gbar^2+\dots)$. Then $\Lambda$ 
is independent of the scale  but dependent on the renormalization
scheme adopted for the coupling. The relation of the 
Schr\"odinger functional $\Lambda$ to its value in other
schemes is given  exactly by a one-loop perturbation theory computation.

For the scale at which we make contact to  physical units we use $L_1$
defined through $\gbar^2(L_1)=4.484$. At this value of the coupling we
have a large and precisely tuned set of pairs $(L_1/a,\beta)$.  
Reanalyzing the data of Ref.~\cite{alpha:nf2} at this point, we 
 get the continuum value of $\Lambda L_1=0.264(15)$.
What remains to be done is to compute $\fk L_1$ in the continuum
limit. To this end, we have to combine the updated data for $L_1/a$ of
\app{a:l1} with the decay constant $\fklat$ at physical quark masses
described in the previous section. Since the data points for $L_1/a$ and
$\fklat$ are at different values of $\beta$, we need to interpolate
$L_1/a$ to the values  of the latter; then the product $\fk
L_1$ can be extrapolated to the continuum limit.

\subsection{Interpolation of $L_1$\label{s:61}}
We start with  values of $L_1/a$ at $\beta=5.2$, $5.2638$, $5.4689$,
$5.619$, $5.758$, $5.9631$ obtained in \app{a:l1} and
\Ref{Blossier:2012qu}.
Since $L_1$ is a physical scale, we expect $\log(L_1/a)$ to be roughly
linear in $\beta$. Four different fits have been tried: a linear fit to the
full range in $\beta$ and to the range [5.2,5.619], which we compared
to quadratic fits in the same intervals. The central values derived from 
these interpolations differ only by a small fraction of the assigned 
statistical uncertainty. Also the statistical errors of the
interpolated points are virtually equal for these procedures, only the
quadratic function to the restricted interval gives larger uncertainties, 
as expected from a fit with three parameters to four points. Since we
want to obtain conservative estimates of the errors, we take the result
of the latter in the further analysis. The three values of $L_1/a$ which
we use in the following are listed in \tab{t:L1inter}.
\begin{table}
\small
\begin{center}
\begin{tabular}{@{\extracolsep{0.2cm}}ccccc}
\toprule
$\beta$  & $L_1/a$ & $L_1\fk$ & $r_0/L_1$ & $\mbar_\strange/\fk$\\
\midrule
$5.2$    &$5.367(82)  $&$0.318(6)(3)$ &$1.155(22)$ &$0.530(12)(6)$\\
$5.3$    &$6.195(51)  $ &$0.320(5)(4)$ &$1.169(15)  $ &$0.577(11)(7)$ \\
$5.5$    &$8.280(80)$ &$0.316(4)(2)$ &$1.213(17)$ &$0.617(11)(5)$ \\
\midrule
cont. & & $0.315(8)(2)$&$1.252(33)$&$0.678(12)(5)$\\
\bottomrule
\end{tabular}
\end{center}
\caption{\label{t:L1inter}\footnotesize Values of $L_1/a$, $L_1
   \fk$, $r_0/L_1$ and
   $\mbar_\strange/\fk$ at the three values of
   $\beta$. For the latter three, we also give the value extrapolated to the
      continuum limit. The running mass in the Schr\"odinger Functional scheme 
      $\mbar_\strange$ is given at the renormalization scale $L_1$.
      Statistical and systematic errors are given.}
\end{table}

\subsection{Continuum value of the $\Lambda$ parameter}
\begin{figure}[tb!]
\vspace{0pt}
\centerline{\includegraphics{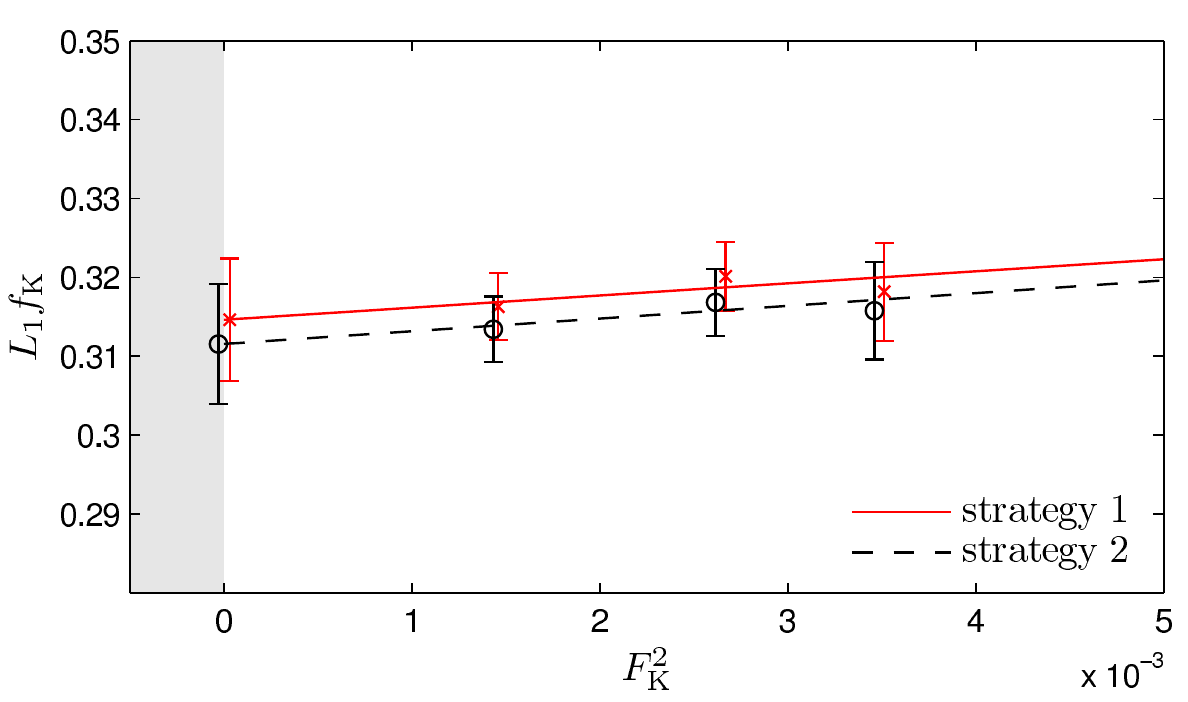}}
  \vspace*{-5mm}
  \caption{\footnotesize
     Continuum extrapolation of $L_1$ in units of $\fk$. Even though the 
        data shows no cut-off effects, we use a linear extrapolation to 
        account for uncertainties from $\rmO(a^2)$ effects hidden
        by the errors. The two strategies for the chiral extrapolation
        of $\fklat=a\fk$ agree well within statistics. 
        }\label{f:Lcont}
\end{figure}

Fig.~\ref{f:Lcont} shows that in the continuum extrapolation of $L_1
\fk$ the cut-off effects are smaller than the statistical uncertainties;
the numerical values are given in \Tab{t:L1inter}.
Indeed, a constant extrapolation gives a $\chi^2/$d.o.f. below unity. However,
we use a linear extrapolation, allowing for $\mathrm{O}(a^2)$ effects
hidden by the statistical fluctuations.\footnote{That the leading corrections
come at $\rmO(a^2)$ is based on the  assumption that the two-loop
approximation of the boundary improvement coefficient $c_t$ of the 
Schr\"odinger functional is sufficient at the accuracy we are aiming at.
This is corroborated by experience from the  quenched 
approximation\cite{pot:intermed} and earlier investigations of the
approach to the continuum with $c_t$ at 1-loop and 2-loop
precision\cite{Sommer:2006sj}.}
In this extrapolation 
the covariance between the three points from the
interpolation in $L_1/a$ is taken into account.  
For comparison with results in the literature,
we give all results also in units of $\left.r_0\right|_{y_1=\ypi}$. The two combinations
evaluate to 
\begin{align}
\fk L_1& = 0.315(8)(2) & r_0/L_1&= 1.252(33)
\end{align}
in the continuum limit.

As in the previous section, the final results come from strategy 1 for
the chiral extrapolation of $\fklat$, given in \tab{t:chiralfits3}. 
Strategy 2 is used to estimate
the systematic uncertainty; however, it is small compared to
the statistical errors and we do not give it separately. 
We therefore quote
\bes
\Lambda^{(2)}/\fk=0.84(6)  \qquad \text{and}\qquad
r_0\,\Lambda^{(2)}=0.331(22)\,.
\ees
Now, as a result of our analysis,
the error is dominated by the error on $\Lambda L_1$.
We translate to the  $\msbar$ scheme using
   $\Lambda_\msbar^{(2)}=2.382035(3)\Lambda^{(2)}$\cite{SF:LNWW,pert:1loop}
   and find
  \bes
\Lambda_\msbar^{(2)} = 310(20)\,\mathrm{MeV}\, \qquad \text{and}\qquad
r_0\,\Lambda_\msbar^{(2)}=0.789(52)\,.
 \ees

\section{Strange quark mass\label{s:7}}
The determination of the strange quark mass uses a
strategy\cite{mbar:pap1} which  splits up the computation into several
steps,  circumventing the multi-scale problem a direct approach would
face. It therefore allows for a good control over the systematic errors. 
Earlier results have been presented in Ref.~\cite{mbar:nf2}, here we
improve on them with a much higher statistical and systematic accuracy.

To briefly summarize the procedure, the RGI mass\footnote{
The RGI mass can be defined through
$M_\mathrm{s}
= \lim_{L \to 0} (2b_0\gbar^2(L))^{-d_0/(2b_0)} \mbar_\strange(L)$ with
the universal coefficient $d_0=8/(4\pi)^2$ \cite{Sint:1998iq}.}
and the renormalized running mass $\mbar_\strange$ in the Schr\"odinger
functional scheme
are given in terms of the bare PCAC mass $m_\strange$ by
\bes
M_\mathrm{s}=Z_\mathrm{M} m_\strange=\frac{M}{\mbar(L)} \mbar_\strange(L)
=\frac{M}{\mbar(L)} \frac{\za}{\zp(L)} m_\strange \,,
\label{e:Ms}
\ees
where the computation of the renormalization constant $Z_\mathrm{M}$ is split
in two steps. The continuum value of the universal first factor 
$M/\mbar$ has been computed in Ref.~\cite{mbar:nf2} for the two 
flavor theory which we consider here. It does not depend on the quark
flavor or the lattice regularization. For the second factor, updated
values of $\zp$ are presented in \app{a:ZP}. 
Of course, the first factor and $\zp$ have to be evaluated at a 
common renormalization scale $L$.  While \Ref{mbar:nf2} used 
$L=\lmax$ with $\gbar^2(\lmax)=4.61$, we choose here $L=L_1$ with 
$\gbar^2(L_1)=4.484$ as in the previous section. The first factor 
in \eq{e:Ms} is $M/\mbar=1.308(16)$ at this scale from a re-evaluation of
the data of \Ref{mbar:nf2}.

\subsection{Determination of the strange quark mass\label{s:kappas1}}

For the bare strange quark mass at physical light quark masses
$m_\strange$, the final ingredient of \eq{e:Ms}, we again follow the two
strategies presented in Sec.~\ref{s:5}.  Of the two,  the second holds
the strange quark mass fixed as a function of the light quark mass and
is therefore the natural choice for this analysis. We still use strategy
1 as a cross check to estimate systematic effects from the chiral
extrapolation.

\subsubsection{Strange mass from strategy 2 \label{s:kappas2} }
The hopping parameter of the strange quark $\kappa_\strange$ is a non-trivial
function of $\kappa_1$, defined in strategy 2 as
\bes
   \kappa_\strange = s(\kappa_1,\mustrange) \,,
\ees
with $s(\kappa_1,\mu)$ defined in \eq{eq:kappas}. 
Its numerical determination
as well as the determination of $\mustrange$
have been explained in \sect{s:heavyK}. 
The resulting values of $\kappa_\strange$ are listed in \tab{tab:ens}.
This definition holds the bare PCAC strange quark mass fixed to the 
value $\mustrange$ set by the physics requirement \eq{eq:mustrange}
at physical light quarks.

Expressing $M_\mathrm{s}$ in units of $\fk$ we eliminate $\za$ and get
\bes
   {M_\strange\over \fkphys} =  {M \over \mbar(L)} \times {1 \over \zp(L)} 
   [1+(\batil-\bptil) \mustrange]\, 
   \frac{ \mustrange}{\fklatphys^\mathrm{bare}}
   \label{e:mstrangergi} \ ,
\ees
with $\fklatphys^{\rm bare} = \fklatphys/\za$.
The second factor is $\rmO(a)$ improved, if we 
neglect, as before, a tiny correction
$(\babar-\bpbar)a \msea$.

\subsubsection{Strategy 1 }

Within this strategy, it is most natural to determine the 
combination $M_\strange+\hat{M}$, the sum of strange and light quark
mass, since this combination is kept
fixed at lowest order in chiral perturbation theory along
the trajectory defined by \eq{e:defhk3}.  
The first order corrections are easily incorporated 
from \cite{PQChPT:Steve} such that we arrive after
renormalization as in \eq{e:mstrangergi} at
\begin{align}
   { 2 m_{13}(\kappa_1,h(\kappa_1)) \over \zp \fk(\kappa_1,h(\kappa_1))}
   =&  {\mbar_\strange +\hat \mbar \over   \fkphys}
       \left[1 + \overline{L}_\mathrm{m}(y_1,\yk) 
        + (\alpha_{4,6} - \frac14)\,(y_1-\ypi) +\rmO(y^2) 
  \right] \label{e:chptm13}
  \\
  \overline{L}_\mathrm{m}(y_1,\yk) =&  L_\mathrm{m}(y_1,\yk) - L_\mathrm{m}(\ypi,\yk) \label{e:lbarm}
  \,,
  \quad \alpha_{4,6} = 3\alpha_4-4\alpha_6 \,,\\
  L_\mathrm{m}(y_1,\yk) =& -(\yk - \frac38 y_1)\log(2\yk/y_1-1) - \yk
\log(y_1) \ ,
  \label{e:lm}
\end{align}
which we can use to extract ${(M_\strange +\hat M) /\fkphys}$ and the 
combination of low energy constants $\alpha_{4,6}$.
Removing the contribution of the average light quark mass $\hat M$ amounts to 
multiplying by a correction factor such that $M_\strange=( M_\strange + \hat M)(1-\rho)$,
where $\rho$ is a small number and can therefore be incorporated 
by its lowest order chiral perturbation theory estimate
\bes
\rho\equiv {\hat M \over M_\strange+\hat M}\approx {\mpi^2 \over
   2 \mk^2} = 0.037 \,.\label{e:rho}
\ees 
Because of this last approximation, we prefer the determination of 
$M_\strange$ from strategy 2 and use the one here just as a 
consistency check. 

\subsection{The strange quark mass in the continuum limit}

\begin{figure}[tb!]
\vspace{0pt}
\centerline{\includegraphics{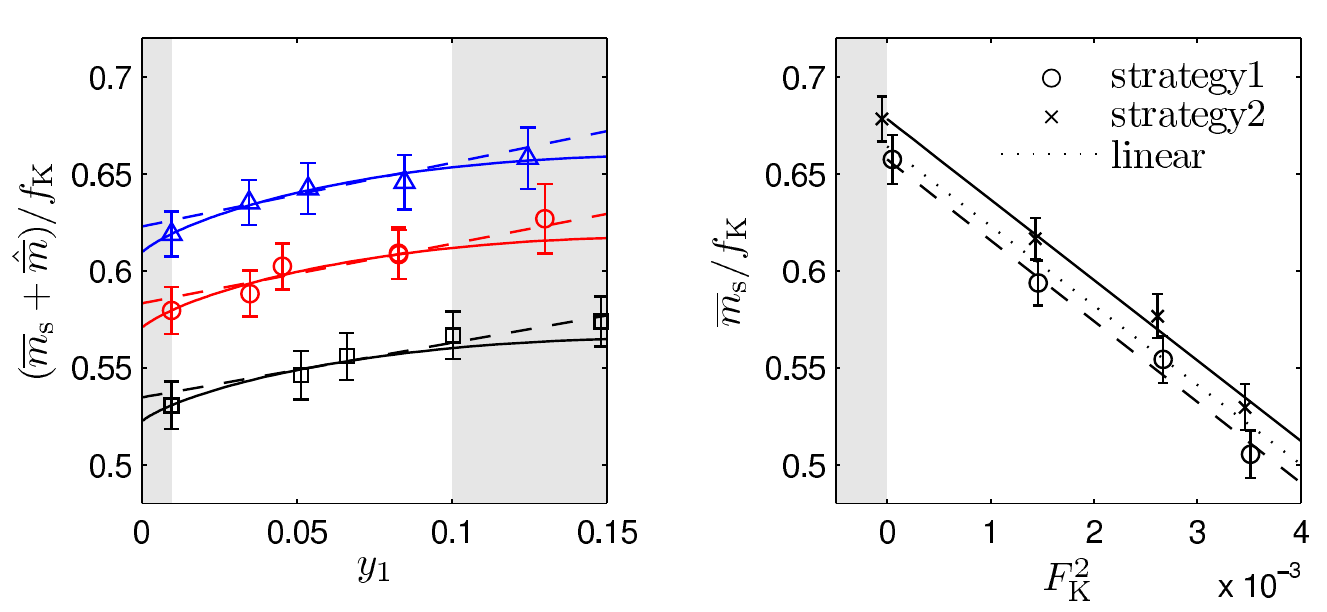}}
  \vspace*{-5mm}
  \caption{\footnotesize
     Left: Chiral extrapolation of the strange quark mass according to 
        strategy 1. The dashed line indicates a corresponding linear fit
        which gives a value that agrees within uncertainties at the
        physical point. Right: Continuum extrapolation of the strange
        quark mass in units of $\fk$ for the results of the two
        strategies.
        \label{f:ms}}
\end{figure}

For strategy 2, the strange quark mass is the strange quark mass is
directly kept fixed at its  physical value. The discussion can be found in
Sec.~\ref{s:5}. For strategy 1, the extrapolation  follows the same
principles as before: we neglect cut-off effects in the NLO terms of the
chiral expansion. The corresponding fit to the data is displayed in
Fig.~\ref{f:ms} on the left, which demonstrates that this
assumption holds well within the statistical accuracy.  Also shown for
comparison is a linear extrapolation which agrees within the
uncertainties with the values obtained from the  ChPT formulae.

The values of $\mbar_\strange/\fk$ for the two strategies as a function
of the lattice spacings are plotted in Fig.~\ref{f:ms} on the right.
Already at finite lattice spacing the two sets seem to be shifted with
respect to another and therefore have parallel continuum extrapolations
which, however,
agree within uncertainties. The data fits well the assumption
that only leading cut-off effects are present and we observe a
correction of 8\% from our value on the finest lattices.
Since we prefer strategy 2 for conceptual reasons, we quote as a final
result with statistical and systematic uncertainties
\begin{align}
\mbar_\strange/\fk&=0.678(12)(5)\,, &M_\mathrm{s}/\fk &= 0.887(19)(7)\,,
   & M_\mathrm{s}&=138(3)(1)\,\mathrm{MeV} \,,
\end{align}
where we use $M/\mbar=1.308(16)$ as quoted before.
For reference, we also give the numbers  in the ${\rm
   \overline{MS\kern-0.05em}\kern0.05em}$  scheme. This 
conversion is the only part of the computation in which we need to take
recourse to perturbation theory, known in this case to four loops, which
differs from the two- and three-loop result by only a small amount. 
We use the same method as described in \Ref{mbar:nf2}, but with the 
new value of $\Lambda_\msbar$ which leads us to
$\mbar^\msbar(2\,\mathrm{GeV})/M=0.740(12)$ and 
\bes
\mbar_\mathrm{s}^\msbar(2\,\mathrm{GeV})=\frac{M_\mathrm{s}}{\fk}\,
\frac{\mbar^\msbar(2\,\mathrm{GeV})}{M}\,  \fkphys = 102(3)(1)\,\mathrm{MeV} \ .
\label{e:msmsbar}
\ees
Here also the statistical
uncertainty of $\Lambda_\msbar$ is included.

\section{Conclusions}
Setting the scale in lattice computations requires a dimensionful
quantity which is known from experiment to good accuracy for which 
we chose the kaon decay constant. 
For currently available lattices this involves chiral extrapolations
to the physical light quark masses, which can be important corrections
depending on the quantity. Chiral perturbation theory turns out
to be a good guide, if one stays within its domain of applicability.
In particular the first strategy discussed
in this paper, designed to have small chiral corrections for the
kaon decay constant, has proven to be successful in the sense that
the extrapolation is very robust under reasonable changes in the 
functional form: various next-to-leading order expansions and also a linear
extrapolation give compatible results.

Whereas strategy 1 keeps the sum of strange and light quark mass 
approximately  constant, strategy 2 provides a constant strange quark
mass, making it suitable for the determination of $M_\strange$ and further
analysis of strange baryons. This flexibility is probably particular
to the situation of a quenched strange quark, however, the fact that
the results from both strategies agree within small uncertainties
gives rise to the expectation that also with a dynamical strange quark
and the u/d quarks sufficiently in the chiral regime, 
the extrapolations can be safe.

As already mentioned in the introduction, this publication 
completes the $\nf=2$ part of a long research program by the ALPHA
collaboration, which has the determination of the $\Lambda$ 
parameter and the strange quark mass as two of its main goals.
We can therefore quote these two observables for the two flavor 
theory in physical units
\begin{align}
\Lambda_{\msbar}^{(2)}&=310(20)\,\mathrm{MeV},&
 M_\strange&=138(3)(1)\,\mathrm{MeV} \,,
\end{align}
with the  value of the strange quark mass in the $\msbar$ scheme 
given in \eq{e:msmsbar}.

Since previous results in physical units were wrong due to inaccurate
values for $r_0/a$ in the literature, we have also  computed 
this scale and get
\bes
  r_0=0.503(10)\,\mathrm{fm} 
\ees
at the physical pion mass using $\fk=155\,\MeV$.

The restriction to two flavor QCD introduces some ambiguity to the scale
setting and quark mass determination, because mass ratios do not have to
be the same as in Nature.  The effect of the heavier quarks is unknown
and interesting in its own right. Comparing to the $\nf=2$ results
presented here, future calculations will shed light on the question of
the contributions of strange and charm quarks in the sea.

\begin{table}
\small
\begin{center}
\begin{tabular}{@{\extracolsep{0.2cm}}cccc}
\toprule
$\nf$  & $\Lambda_\msbar$ & experiment & theory\\
\midrule
$0$      & $238(19)\,\MeV$  & $m_\mathrm{K}$, $K\to \mu\nu_\mu$, $K\to
\pi\mu\nu_\mu$ & lattice gauge theory\cite{alpha:SU3}\\
$2$      & $310(20)\,\MeV$  & $m_\mathrm{K}$, $K\to \mu\nu_\mu$, $K\to \pi\mu\nu_\mu$ 
& this work\\
\midrule
$5$      & $212(12)\,\MeV$      & world average & perturb.
theory\cite{Bethke:2011tr}\\
\bottomrule
\end{tabular}
\end{center}
\caption{\footnotesize\label{t:comp}
   The $\Lambda$-parameter as a function of the number of
      flavors.
  }
\end{table}

For now we can make a comparison to the values obtained
in the quenched approximation. In the strange quark
mass we find no significant difference to the corresponding value quoted in
\Ref{mbar:pap3}. However, the $\Lambda$ parameter exhibits
an interesting dependence on $\nf$, with 
the five flavor theory $\Lambda^{(5)}=212(12)\,\MeV$ 
extracted from Bethke's  world average 
$\alpha_\msbar(M_{\rm Z})=0.118(10)$~\cite{Bethke:2011tr} and 
the quenched value $\Lambda^{(0)}=238(19)\,\MeV$\cite{alpha:SU3} listed in \Tab{t:comp}.
 Perturbation theory predicts
$\Lambda^{(\nf)}<\Lambda^{(\nf-1)}$, which is valid and precise 
when the quark being decoupled is sufficiently heavy. 
This is expected to be the case for the b-quark and thus 
$\Lambda^{(5)} < \Lambda^{(4)}$. However, comparing the 
ALPHA value for $\nf=0$ and our new $\nf=2$ result  shows that for 
light quarks the $\nf$-dependence is in the opposite direction. 
The $\Lambda$ parameter increases quite strongly with increasing 
$\nf$. An eventual decrease is likely to be present between 
$\nf=3$ and $\nf=4$, leading to an  agreement with the 
phenomenological determinations of 
$\alpha_\msbar(M_{\rm Z})$~\cite{Bethke:2011tr}.

In any case, it will be an important milestone to arrive at a
non-perturbative and precise determination of $\Lambda^{(4)}$ and
compare with the present world average. This is particularly needed in
the light of the spread of different determinations before taking the
world average.  For example the authors of  \Ref{Alekhin:2012ig} analyse
deep inelastic scattering data by a global fit to parton distribution
functions and $\alpha_\msbar$ at NNLO. Their value for
$\alpha_\msbar(M_{\rm Z})$ corresponds to
$\Lambda^{(5)}_\msbar=160(11)\,\MeV$, which is  quite a bit smaller than
the world average.

\vspace{0.5cm}

\noindent
{\bf Acknowledgements.}
We acknowledge helpful  discussions with 
O.~B\"ar, M.~Della Morte, M.~L\"uscher, H.~Simma, U.~Wolff 
and in general members of CLS. We are grateful for
computer time allocated for our project  on the Jugene and Juropa
computers at NIC, J\"ulich, 
and the ICE at ZiB, Berlin. The lattice group at the University of Bielefeld
kindly granted us access to their apeNEXT computers, which allowed
us to improve the determination of the renormalization constants.
This work is supported by the Deutsche Forschungsgemeinschaft
in the SFB/TR~09 and the GRK 1504 ``Masse, Spektrum,
   Symmetrie'' as well as by the European community
through EU Contract No.~MRTN-CT-2006-035482, ``FLAVIAnet''.
This work was granted access to the HPC resources of the
Gauss Center for Supercomputing at Forschungzentrum J\"ulich, Germany,
made available within the Distributed
European Computing Initiative by the PRACE-2IP, receiving funding from the
European Community's Seventh Framework Programme (FP7/2007-2013) under grant
agreement RI-283493.

\begin{appendix}

\section{Details of the update algorithms\label{a:1}}
In our simulation we first used the DD-HMC
algorithm\cite{algo:L2,algo:L3}
as implemented in L\"uscher's package\cite{soft:DDHMC} and then switched
to our implementation of a mass preconditioned
HMC\cite{Marinkovic:2010eg}.
Both algorithms have several parameters which influence their
performance and we therefore give the essential ones in this
section.

\subsection{DD-HMC}
The block decomposition of the DD-HMC algorithm separates
the infrared from the ultraviolet modes. The details and notation can be found
in Ref.~\cite{algo:L2}. We give the size of the blocks and the step
sizes of the three different forces in Table~\ref{tab:ddhmc}. The 
gauge force has $N_0 N_1 N_2$ steps per trajectory, the block
fermion force $N_1 N_2$ and the global fermion force $N_2$ steps.
The number of trajectories times their length is given in the ``MD
time'' column of the table; to get an effective statistics, this can be
multiplied by the ratio of active links $R_\mathrm{active}$.

\begin{table}
\small
\begin{center}
\begin{tabular}{lcccccccc}
\toprule
id & block & $R_\mathrm{active}$    &  $\tau$ &$N_0$ & $N_1$ & $N_2$ & acc. rate & MD time  \\
\midrule
A3 & $8^4$ & $0.37$ & $2$       & $4$    & $5$     & $125$   & $0.91$      & $8030$\\  
A4 & $8^4$ & $0.37$ & $2$       & $4$    & $5$     & $150$   & $0.85$      & $8090$\\  
\midrule
E5f& $8^4$ & $0.37$ &  $0.5$    & $4$    & $5$     & $22$    & $0.87$      & $16000$\\
E5g& $8^4$ & $0.37$ &  $4$      & $4$    & $5$     & $176$   & $0.83$      & $16180$\\
F6 & $8^4$ & $0.37$ &  $2$      & $4$    & $5$     & $260$   & $0.89$      & $4800$\\
F7a& $8^4$ & $0.37$ &  $2$      & $4$    & $5$     & $350$   & $0.84$      & $5600$\\
F7b& $8^4$ & $0.37$ &  $2$      & $4$    & $5$     & $350$   & $0.87$      & $4000$\\
\midrule
N4 & $8^2 12^2$& $0.44$ & $0.5$ &$4$     & $5$     & $24$    & $0.88$ &$3700$\\
N5 & $8^2 12^2$& $0.44$ & $0.5$ &$4$     & $5$     & $24$    & $0.87$ &$3800$\\
\bottomrule
\end{tabular}
\end{center}
\caption{\footnotesize\label{tab:ddhmc}Parameters of the DD-HMC algorithm for the
ensembles used in this publication. We give the HMC block size, the
corresponding ratio of active links $R_\mathrm{active}$, the trajectory length, the
(relative) step sizes of the gauge, block fermion and global fermion
force. This is followed by the measured acceptance rate and the total
statistics after thermalization.}
\end{table}

\subsection{MP-HMC}
In our implementation of the mass preconditioned
HMC\cite{Marinkovic:2010eg}, we split the fermion determinant $\det Q^2$,
with $Q=a\gamma_5D$,
first by symmetric even-odd preconditioning 
\be
\det Q^2 ={\det}^2 Q_\mathrm{ee} \,  {\det}^2 Q_\mathrm{oo}  \, {\det}^2 Q_\mathrm{S}
\ee
and then apply two levels of 
mass preconditioning\cite{algo:GHMC} to its Schur complement
$Q_\mathrm{S}=1-Q_\mathrm{ee}^{-1} Q_\mathrm{eo} Q_\mathrm{oo}^{-1}
Q_\mathrm{oe}$\,, i.e.,
\be
\det Q_\mathrm{S}^2 = 
           {\det}^2 \big [ W^{-1}( \sigma_2) W( \sigma_1)\big ]\,
           {\det}^2 \big [ W^{-1}( \sigma_1) Q_\mathrm{S}\big ]\,
           {\det}^2 \big [ W( \sigma_2) ]
\label{eq:mp}
\ee
with $W(\sigma)=Q_\mathrm{S}+\sigma$ and $\sigma_2>\sigma_1>0$. The forces from the 
individual contributions to the action are integrated on 
multiple time scales\cite{Sexton:1992nu}, the gauge force is on the finest level,
the determinants of $Q_\mathrm{ee}$ and $Q_\mathrm{oo}$
are integrated together with the third determinant of \eq{eq:mp},
the next two determinants constitute the two coarsest levels.  For
each we use a second order integrator\cite{Omelyan2003272},
where an elementary step of size $\epsilon$ is given by
\be
T(\epsilon)=T_1(\lambda \epsilon)\, T_2(\epsilon/2)\,
   T_1((1-2\lambda) \epsilon)\,
T_2(\epsilon/2)\, T_1(\lambda \epsilon) \ .
\label{eq:om}
\ee
$T_1$ is the force application on each level, whereas $T_2$ is again
given by such a step with the force of the lower level. We use
$\lambda=0.19$ for all levels. Only the gauge force is integrated with a
standard leapfrog. Analogously to the DD-HMC, we give the relative
step sizes in Table~\ref{tab:mphmc}, note however, that \eq{eq:om}
implies a factor of two in the number of force applications between
levels even for $N_i=1$. The preconditioning masses are given in
terms of the hopping parameter such that
$2\sigma_i=\kappa_i^{-1}-\kappa_\mathrm{sea}^{-1}$.

\begin{table}[tb!]
\small
\begin{center}
\begin{tabular}{lccccccccc}
\toprule
id & $\kappa_1$ & $\kappa_2$ & $\tau$ & $N_0$ &  $N_1$ & $N_2$ & $N_3$ & acc. rate & MD time \\
\midrule
A5c & 0.135887       & 0.134250         & 2      & 9     &    1   &  1    & 32    & 0.93 &1160\\
A5d & 0.135887       & 0.134250         & 2      & 9     &    1   &  1    & 32    & 0.92 &1700\\ 
\midrule
N6  & 0.136552       &  0.133857       & 2      & 9     &    1    & 1     &16    & 0.84& 4000\\
O7  & 0.136550       &  0.135000       & 2      & 9     &    1    & 2     &16    & 0.83& 4000\\
\bottomrule
\end{tabular}
\end{center}
\caption{\footnotesize\label{tab:mphmc}
      Parameters of the MP-HMC algorithm for the ensembles used in this
      publication. The $\kappa_i$ parametrize the preconditioning masses
      and are followed by
      the trajectory length and the (relative) step numbers per
      trajectory. Also the acceptance rate and the statistics after
      thermalization are given.}
\end{table}

\section{Improved determination of $\za$\label{a:ZA}}

The renormalization factor $\za$ was previously 
computed~\cite{DellaMorte:2008xb} from a carefully chosen Ward identity.
At $\beta=5.2$ an error of  $2.1\%$ resulted from the fact that
at this $\beta$ an extrapolation in the quark mass was necessary.
Since this uncertainty directly propagates into the strange quark mass 
determination and
scale setting, we here improve $\za$ at $\beta=5.2$.
We substantially increased the statistics for the three previously
simulated quark mass points and added another 
mass point,
in order to better control the subsequent chiral extrapolation.
We follow the lines of~\cite{DellaMorte:2008xb}, i.e., 
in the determination of 
$\za$ we use the optimal wave function of~\cite{impr:ca_nf2} to suppress the
first excited state contribution in the pseudoscalar channel, 
use the ``massive''
definition of the renormalization constant~\cite{impr:za_nf2}, and we drop the
disconnected quark diagrams which only amount to 
${\rm O}(a^2)$ effects on $\za$.
In the notation of~\cite{DellaMorte:2008xb} we thus compute $\za^{\rm con}$ with
$\omega=\omega_{\pi^{(0)}}$.
The results are listed in \tab{tab:ZA-beta52}.
\begin{table}
\small
        \centering
        \begin{tabular}{ccccccl}
        \toprule
         $\beta$  & $L/a$ & $T/a$ & $\kappa$   &  $\za$      &  $am$          \\\midrule
          $5.20$  & $12$  & $18$  & $0.13550$  & $0.7853(15)$&  $0.02082(20)$ \\[0.2em]
                  &       &       & $0.13560$  & $0.7792(20)$&  $0.01700(20)$ \\[0.2em]  
                  &       &       & $0.13570$  & $0.7767(20)$&  $0.01345(22)$ \\[0.2em] 
                  &       &       & $0.13580$  & $0.7757(27)$&  $0.00892(15)$ \\\cmidrule(lr){4-6}
                  &       &       & ---        & $0.771(6)$ &  $\to 0$           \\\bottomrule
        \end{tabular}
        \caption{\footnotesize Simulation parameters and results for $\za$ and the PCAC mass $am$ at $\beta=5.2$.
                 The last row shows the chirally extrapolated value as described in the main text.
                 To be compared with Table~1 of~\cite{DellaMorte:2008xb} and
                 its chiral value $ 0.774(16)$.
                 }
        \label{tab:ZA-beta52}
\end{table}
\begin{figure}
        \begin{center}
            \includegraphics{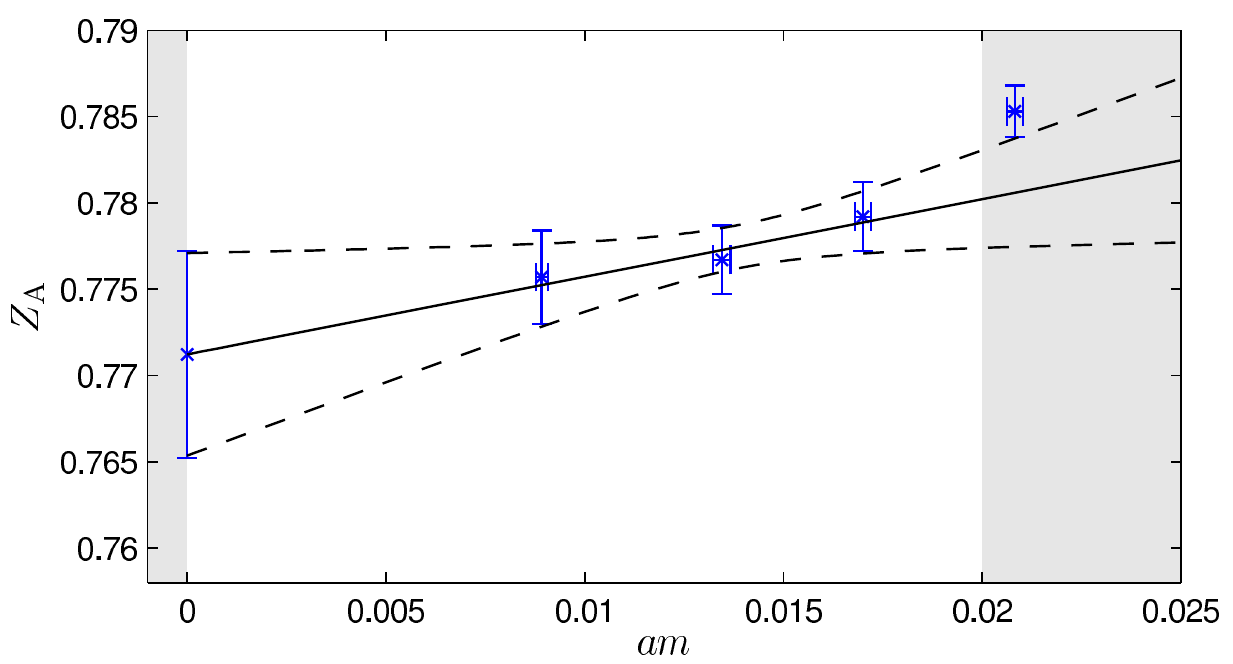}
        \end{center}
        \caption{\footnotesize Data points with highly improved statistics for $\za$ at $\beta=5.2$.
                 The point in the chiral limit is obtained by a linear 
                 extrapolation of the three lightest PCAC masses.
                 }
        \label{fig:ZAcon-fit-beta52}
\end{figure}

\begin{figure}
        \vskip1em
        \begin{center}
            \includegraphics{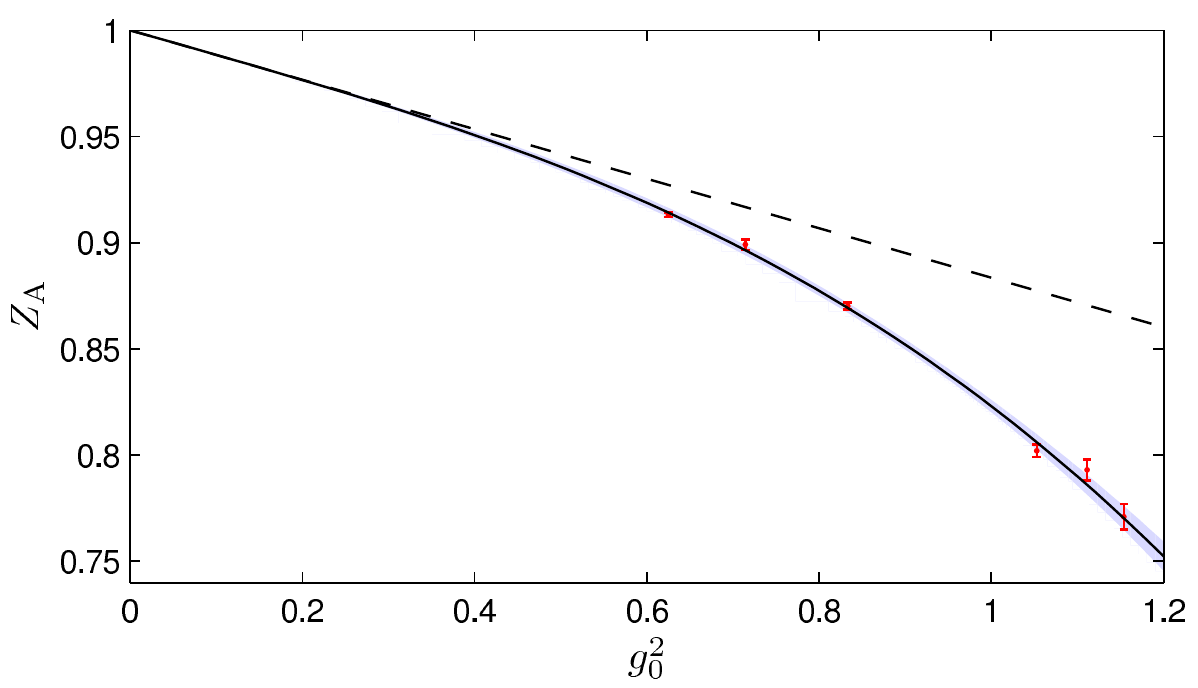}
        \end{center}
        \caption{\footnotesize Result for our global fit of $\za\big(g_0^2\big)$, 
                 together with the error band from eq.~\eqref{eq:ZA-result}. For comparison
                 we also show the perturbative (1-loop) result for
                 $\nf=2$.}
        \label{fig:ZA-gloabl-fit}
\end{figure}

We have performed several chiral extrapolations, including weighted averages of the two
and three data points with the smallest masses, but in the end all resulting central
values at $am=0$ are covered by our preferred ansatz, a linear extrapolation of 
the data points with $am<0.02$. It is shown in \fig{fig:ZAcon-fit-beta52}. The total
range of the $y$-axis of the figure is the previous~\cite{DellaMorte:2008xb} 
$\pm1\sigma$ error. We now have a considerably improved value of $\za=0.771(6)$. 
What remains to be done is to provide a global fit for $\za(g_0^2)$.
Again, several fit ansaetze have been tried and all of them give comparable results,
especially in the $\beta$-range which is important in the present paper. 
Constraining the asymptotic behaviour to the perturbatively known one-loop estimate,
we finally obtain
\begin{align}\label{eq:ZA-result}
  \za(g_0^2) &= 1-0.116458\,g_0^2 +c_1\,g_0^4  +c_2\,g_0^6 \;, \\[0.5em] \notag
      c &=
    \begin{pmatrix}
    +1.16   \\
    -7.21   \\
    \end{pmatrix} \cdot 10^{-2} \;, &
   \hspace*{-2cm}{\rm Cov}(c) &= 3\cdot10^{-4} \times
    \begin{pmatrix}
    +0.74 & -0.75  \\
    -0.75 & +0.79  \\
    \end{pmatrix} \;,
\end{align}
where ${\rm Cov}(c)/3$ is the covariance matrix of the parameters $c_1,c_2$ obtained
directly from the statistical errors of the data points.
Due to the constraint fit ansatz with just two parameters, the error of the 
fit is significantly smaller than that of the individual data points.  
We therefore recommend 
to conservatively use ${\rm Cov}(c)$ with the factor 3 applied as above,
which yields the grey error band in \fig{fig:ZA-gloabl-fit}.

\section{Determination of $\zp$\label{a:ZP}}
From the simulations reported in~\cite{Blossier:2012qu} we also have
values for $\zp$, the renormalization constant of the pseudoscalar
density. They are based on simulations with 8000 essentially uncorrelated measurements
and thus have a very small statistical uncertainty. However, the small
mismatch of the simulated $(L/a,\beta,am)$ has to be accounted for
and introduces additional errors. We need to distinguish
$\beta=5.2$ and $\beta\geq 5.2638$, since in the latter case, 
we only have very small shifts.

We start with $\beta=5.2$. 
Here we have two different sea quark masses each
at $L/a=4,6$ (see also \tab{tab:gbsq-L-to-L1}). 
They enclose $Lm=0$. We thus first perform an interpolation to vanishing mass and 
subsequently interpolate $\ln \zp$ to our target coupling linearly in $\ln\gbar^2$, 
c.f. Ref.~\cite{mbar:nf2}. A systematic error
is estimated by comparing to an interpolation linear in $\gbar^2$.

Now we turn to $\beta\geq 5.2638$. Both the shifts to $am=0$,~$\gbar^2=4.484$
and the propagation of the errors in $\gbar^2(L)$ of \tab{tab:gbsq-L-to-L1}
is done by a first order expansion of 
\begin{align}
        F(x,y) &= \left.\ln\zp\right|_{\beta=\mathrm{const}}
        && \text{with } x=\ln(\gbar^2),\ 
                  y=Lm\,,
        \label{eqn:taylor-lnZP}
\end{align}
in $x,y$. The derivative $\partial_x F$
is computed from perturbation theory, using the 
known (continuum) two-loop and three-loop perturbative results 
for $\tau(\gbar)$ and $\beta(\gbar)$, respectively. Note that
perturbation theory is known to be rather accurate for 
the running of the mass (and therefore $\zp$) in the SF-scheme~\cite{mbar:nf2}.
The other derivative is estimated as follows. At $\beta=5.2$, $L/a=4,6$ we have 
two mass points each and estimate $\partial_y F$ for $\gbar^2=3.80,\,4.95$
by finite differences. These
are then interpolated to
$\partial_y F(\ln(4.484),0)=0.148(30)$. Up to tiny effects coming from
the renormalization of $m$, the derivative $\partial_y F$ is universal.
We neglect the $a^2/L^2$ corrections to exact universality,
using  $\partial_y F(\ln(4.484),0)=0.148(30)$ at all $\beta$. 
 
The results for $\zp$ are collected in \tab{tab:zp-summary}.  
Apart from the special case $\beta=5.2$, where the 
interpolation in $\gbar^2$ introduces a noticeable uncertainty,
all errors in the last column are dominated by the statistical 
errors in $\gbar^2(L_1)$. 
\begin{table}
\small
\begin{center}
\begin{tabular}{ccccccccc}\toprule
   $L/a$ & $\beta$  & $\kappa$   & $am$             & $\zp(L,\beta,am)$ & $\zp(L_1,\beta,0)$ \\\midrule
   $4$   & $5.2$    & $0.135041$ & $-0.00279(14)$   &  $0.57672(15)$    &   --               \\
         &          & $0.134800$ & $+0.00464(20)$   &  $0.58291(22)$    &   --               \\
         &          &            & $\to 0$          &  $0.57904(13)$    &   --               \\
   $6$   & $5.2$    & $0.135617$ & $-0.00747(20)$   &  $0.48796(28)$    &   --               \\
         &          & $0.135226$ & $+0.00571(17)$   &  $0.49641(24)$    &   --               \\
         &          &            & $\to 0$          &  $0.49275(18)$    &   --               \\\cmidrule(lr){1-6}
         & $5.2$    &            & $0$              &  --               &   $0.5237(45)$     \\
   $  6$ & $5.2638$ & $0.135985$ & $-0.00585(18)$   & $0.50676(28)$     &   $0.5150(16)$     \\
   $  8$ & $5.4689$ & $0.136700$ & $-0.00339(13)$   & $0.51103(29)$     &   $0.5170(17)$     \\
   $ 10$ & $5.6190$ & $0.136785$ & $-0.00260(9)\;\,$& $0.51255(30)$     &   $0.5186(25)$     \\
   $ 12$ & $5.7580$ & $0.136623$ & $+0.00040(6)\;\,$& $0.51856(29)$     &   $0.5187(43)$     \\
   $ 16$ & $5.9631$ & $0.136422$ & $-0.00107(4)$   & $0.51879(35)$     &   $0.5172(48)$     \\\bottomrule
\end{tabular}
        \caption{\footnotesize Pseudoscalar renormalization constant $\zp$ from simulations in the 
                 Schr\"odinger functional. The kinematical setup is different from those
                 in \tab{tab:gbsq-L-to-L1} and thus the PCAC mass $am$ differs.
                 The rightmost column lists the results after correcting for 
                 an imperfect tuning as explained in the text. For
                 $\beta=5.2$, two non-zero quark masses have been
                 simulated, allowing for an interpolation to vanishing
                 mass.
                 }
        \label{tab:zp-summary}
\end{center}
\end{table}

Within our uncertainties the $g_0^2$-dependence of $\zp$ 
at fixed $L=L_1$ is not clearly visible. We thus take the average value of all six results 
and assign an error of about $1\%$ to cover all results: 
\begin{align}
        \zp &= 0.5184(53) \;,  &
        5.2 \leq \beta \leq 6.0  \;.
        \label{eqn:ZP-final}
\end{align}

\section{Improved determination of  $L_1$\label{a:l1}}

The practical reason for $\gbar^2(L_1)=4.484$ instead of 
 $\gbar^2(L_{\rm max})=4.61$, chosen in \cite{alpha:nf2}, 
is that meanwhile
dedicated tuning runs were carried out~\cite{Blossier:2012qu},
which fix the simulation parameters 
$(L/a,\beta,\kappa_{\rm sea})$ at the line of constant physics
\begin{align}
        \gbar^2(L_1) &= 4.484  \;,  &
                L_1m   &= 0  \;.
        \label{eq:locp-L1}
\end{align}
Here $m$ is the PCAC mass of the degenerate sea quarks and additional 
details can be found in the appendices of~\cite{Blossier:2012qu}.
Of course fixing such a line of constant physics is only possible within
a certain (statistical) uncertainty.
While in the results of~\cite{Blossier:2012qu} the error 
due to the line of constant physics is not so relevant, we here aim for a more precise
setting in order to extract the $\Lambda$ parameter and the scale $L_1$ in 
physical units. For this reason we have set up new simulations for $\gbar^2$
at $\beta\in\{5.2,5.2638,5.4689,5.6190\}$, corresponding to the first five 
sets of parameters in \tab{tab:gbsq-L-to-L1} which enclose the $\beta$-values 
used here. For completeness we also list the additional two sets for
$\beta\in\{5.7580,5.9631\}$, taken over from the aforementioned reference.
The simulations for $\gbar^2$ proceed along the lines of~\cite{alpha:nf2}.
Results are listed in \tab{tab:gbsq-L-to-L1} together with the deviation
$\delta[\gbar^2]=\gbar^2(L)-4.484$. 
At $\beta=5.2$ there is a large deviation since  our code does not allow to 
simulate $L/a=5$; we have to rely on neighbouring points with $L/a=4,6$.
At each of these values of $L/a$ we have simulations at
slightly positive and negative $m$ from which we interpolate to
$Lm=0$, see \tab{tab:gbsq-L-to-L1}.
We finally just take the value from $L/a=6$ into account, since it is closer 
to our target $\gbar^2$.
At fixed lattice spacing we now take the results of $\gbar^2(L)$ 
as computed at integer $L/a$ and correct for the desired value $L_1/a$ at
$\gbar^2(L_1)=4.484$, using 
\begin{align}
  \ln\left[\frac{L_1}{a}\right] &= 
  \ln\left[\frac{L}{a}\right]+\int_{\gbar(L_1)}^{\gbar(L)}\!\frac{{\rm d}x}{\beta(x)} \;.
   \label{eq:RG-gbsq-vs-Loa}
\end{align}
The non-perturbative estimate of the $\beta$-function
from~\cite{alpha:nf2} is inserted, but we treat the difference to the value
of $L_1/a$, obtained from perturbation theory at the highest available order, as 
a systematic uncertainty.\footnote{
The estimate of $\beta(x)$ in~\cite{alpha:nf2} applies to the continuum 
step scaling function, while we here apply it for not-so-large $L/a$. 
This represents a second reason to prefer $L/a=6$ to $L/a=4$ in our determination
at $\beta=5.2$.
}
 We quote this as second error in our results for $L_1/a$ in 
\tab{tab:gbsq-L-to-L1},
the first error is due to the statistical error of $\gbar^2(L)$. 
Our final numbers to be used in Section~\ref{s:61} are those shown in the last column of
\tab{tab:gbsq-L-to-L1}.

\begin{table}
\small
\centering
\begin{tabular}{cccccccccc}
\toprule
      $L/a$ & $\beta$  & $\kappa_{\rm sea}$ & $am$            & $\gbar^2(L)$  & $\delta[\gbar^2]$  &   $L_1/a$ \\\midrule
        $4$ & $5.2000$ &  $0.134700$        & $-0.03745(41)$  & $3.730(11)$    \\
        $4$ & $5.2000$ &  $0.133780$        & $-0.00086(35)$  & $3.797(11)$    \\
        $4$ & $5.2000$ &  --                & $\to 0$         & $3.798(11)$ &   $-0.686$   &   $  5.11(3)(13)$    \\
        $6$ & $5.2000$ &  $0.135600$        & $-0.01322(26)$  & $4.810(32)$    \\
        $6$ & $5.2000$ &  $0.135200$        & $+0.00289(24)$  & $4.984(33)$    \\
        $6$ & $5.2000$ &  --                & $\to 0$         & $4.954(33)$ &   $+0.470$   &   $  5.33(4)(11)$    \\\cmidrule(lr){1-6}
        $6$ & $5.2638$ &  $0.135673$        & $+0.00012(19)$  & $4.550(25)$ &   $+0.066$   &   $  5.89(4)(2)$     \\
        $8$ & $5.4689$ &  $0.136575$        & $+0.00046(11)$  & $4.526(32)$ &   $+0.042$   &   $  7.91(7)(1)$     \\
       $10$ & $5.6190$ &  $0.136700$        & $+0.00038(8)$   & $4.531(51)$ &   $+0.037$   &   $  9.87(14)(2)$    \\\cmidrule(lr){1-6}\addlinespace[0.005cm]\cmidrule(lr){1-6}
       $12$ & $5.7580$ &  $0.136623$        & $+0.00067(7)$   & $4.501(91)$ &   $+0.017$   &   $ 11.94(31)(1)$    \\
       $16$ & $5.9631$ &  $0.136422$        & $-0.00096(4)$   & $4.40(10)$  &   $+0.084$   &   $ 16.40(50)(6)$    \\
\bottomrule
\end{tabular}
\caption{\footnotesize Values of $L_1/a$ after correcting the simulated values $L/a$. For completeness we also
         include the two largest $\beta$-values, where no new simulations enter. The second error
         on the final result is the systematic one.}
\label{tab:gbsq-L-to-L1}
\end{table}

\section{\label{a:mcrit}Critical mass $m_\mathrm{cr}$}

Since Wilson fermions explicitly break chiral symmetry, the quark masses
experience an additive renormalization $m_\mathrm{cr}$. It is defined by
the point where the PCAC quark mass \eq{eq:m} vanishes for
quarks degenerate to the sea quarks: $m_{12}=0$.  Even though we have chosen our
improvement procedure in terms of $m_{rs}$ and hence do not
need $m_\mathrm{cr}$ in our analysis, we give here its determination
for completeness. 

Up to terms of order $a^2$ the subtracted bare quark mass 
$m_\mathrm{q}=m_0-m_\mathrm{cr}$ is proportional to the quark mass from the 
plateau of \eq{eq:m}
\be
m_{12}(1+\tilde b_\mathrm{r} a m_{12}) 
= Z_\mathrm{m}\frac{\zp}{\za}r_\mathrm{m}\,  m_\mathrm{q} \equiv Z\, r_\mathrm{m}\, m_\mathrm{q}
\, , 
\label{eq:mcrit}
\ee
which we use to define the subtraction term $m_\mathrm{cr}$.
The improvement coefficient $\tilde b_\mathrm{r}$ is a linear combination
of those improving the axial current, the pseudoscalar density and the
bare subtracted quark mass while $Z_\mathrm{m}$ renormalizes quark mass
differences $m_{\mathrm{q},i} - m_{\mathrm{q},j}$ and $Z_\mathrm{m}r_\mathrm{m}$
renormalizes the trace of the bare subtracted mass matrix $\sum_i
m_{\mathrm{q},i}$\cite{impr:nondeg}.
In one-loop perturbation theory,
$ \tilde b_\mathrm{r}=\frac{1}{2}+0.0500\,g_0^2$\cite{impr:pap5,impr:babp}, 
which we expect to be a
sufficient approximation since our quark masses are small.
Results are given in \tab{tab:mcrit}, where we quote 
$\kappa_\mathrm{cr}=(2am_\mathrm{cr}+8)^{-1}$ instead of $m_\mathrm{cr}$
and use the non-perturbative
$
  Z =1+0.090514\,g_0^2\, (1-0.3922\,g_0^4-0.2145\,g_0^6)/(1-0.6186\,g_0^4)
$
of \cite{impr:babp_nf2} to determine $r_\mathrm{m}$ from $Zr_\mathrm{m}$.
The central values
are from fits of \eq{eq:mcrit} to the data with $am_{12}<0.01$; the
resulting $\chi^2/$d.o.f. are all smaller than 1, which might
be an indication that the errors on $m$ are overestimated. The systematic
error is estimated from including also a free cubic term on the right
hand side of \eq{eq:mcrit} and a variation of the fit range up to
$am_{12}=0.016$.

\begin{table}
\small
\begin{center}
\begin{tabular}{@{\extracolsep{0.2cm}}cccc}
\toprule
$\beta$ & $\kappa_\mathrm{cr}$ & $Zr_\mathrm{m}$ & $r_\mathrm{m}$\\
\midrule 
5.2&0.1360546(25)(30) & 1.438(20)(30)  & 1.549(42)\\
5.3&0.1364572(11)(30) & 1.310(11)(20)  & 1.323(23)\\
5.5&0.1367749(06)(04)  & 1.228(06)(05) & 1.157(08)\\
\bottomrule

\end{tabular}
\caption{\label{tab:mcrit} \footnotesize Value of the critical hopping parameter and
the renormalization constant $Zr_\mathrm{m}$ and $r_\mathrm{m}$ for our three values of
the coupling constant. The first error in $Zr_\mathrm{m}$ is statistical, the second
systematic from a variation of the functional form and the range of the
fit to \eq{eq:mcrit}. For $r_\mathrm{m}$ these errors are added in quadrature
together with the one from $Z$.}
\end{center}
\end{table}

These non-perturbative values for $Zr_\mathrm{m}$ and $r_\mathrm{m}$ deviate significantly from the
one-loop expectation $Zr_\mathrm{m}=1+0.090514\,g_0^2$\cite{impr:babp}, which is
roughly $Zr_\mathrm{m}\simeq 1.1$ in our region of couplings and even more so 
for $r_\mathrm{m}$ which is one up to $\rmO(g_0^4)$ corrections.

\section{Masses of the pseudoscalar mesons, decay constants and quark
masses\label{app:a2}}
Following the analysis described in \sect{s:dana}, we determined starting time 
slices $x_0^{\mathrm{min}}/a$ whose value for the ensembles we give in
\Tab{tab:x0min}. We give one value each for the mesons composed of
two sea quarks, a sea and a strange quark and two strange quarks.
We list the raw data of quark masses, pseudoscalar masses
and decay constants in \Tabs{tab:mpsfps52}--\ref{tab:mpsfps55}.

\begin{table}[ht]
\small
\begin{center}
\begin{tabular}{@{\extracolsep{0.1cm}}lcccccccccccc}
\toprule
                         & A2 & A3 & A4 & A5 & E4 & E5 & F6 & F7 & N4 & N5 &  N6 & O7\\
\midrule
($\kappa_1$, $\kappa_2$) &$17$&$17$&$17$&$13$&$15$&$20$&$17$&$17$&$26$&$23$& $23$&$23$\\
($\kappa_1$, $\kappa_3$) &$18$&$18$&$18$&$15$&$16$&$22$&$22$&$21$&$27$&$27$& $28$&$28$\\
($\kappa_3$, $\kappa_4$) &$19$&$19$&$19$&$19$&$18$&$24$&$24$&$26$&$23$&$29$& $31$&$31$\\
\bottomrule
\end{tabular}
\end{center}
\caption{\label{tab:x0min} \footnotesize Values of $x_{0}^\mathrm{min}/a$ used
 in the  analysis, from top to bottom, for mesons made of two sea quarks,
 a sea and a strange quark and two strange quarks.}
\end{table}

\begin{table}[p!]
\small
\begin{center}
\begin{tabular}{@{\extracolsep{0.2cm}}lllllll}
\toprule
 &  $\kappa_r$ & $\kappa_s$ & $am_{rs}$ & $am_\mathrm{PS}$ &
 $F_\mathrm{PS}^\mathrm{bare}$  \\
\midrule
\multirow{8}{*}{A2} 
&$0.13510$ &$0.13510$  & $0.02934(5)$ & $0.3250(5)$ & $0.1009(2)$\\
&$0.13530$ &$0.13530$  & $0.02438(5)$ & $0.2968(5)$ & $0.0977(2)$\\
&$0.13550$ &$0.13550$  & $0.01946(5)$ & $0.2661(5)$ & $0.0943(2)$\\
&$0.13570$ &$0.13570$  & $0.01454(5)$ & $0.2315(6)$ & $0.0908(2)$\\
&$0.13565$ &$0.13510$  & $0.02252(5)$ & $0.2858(5)$ & $0.0963(2)$\\
&          &$0.13530$  & $0.02007(5)$ & $0.2702(5)$ & $0.0947(2)$\\
&          &$0.13550$  & $0.01762(5)$ & $0.2537(6)$ & $0.0930(2)$\\
&          &$0.13565$  & $0.01578(5)$ & $0.2407(6)$ & $0.0917(2)$\\
\midrule
\multirow{8}{*}{A3} 
&$0.13510$ &$0.13510$  & $0.02707(5)$ & $0.3078(4)$ & $0.0965(2)$\\
&$0.13530$ &$0.13530$  & $0.02211(5)$ & $0.2785(5)$ & $0.0930(2)$\\
&$0.13550$ &$0.13550$  & $0.01720(5)$ & $0.2464(5)$ & $0.0893(2)$\\
&$0.13570$ &$0.13570$  & $0.01231(6)$ & $0.2096(6)$ & $0.0854(2)$\\
&$0.13580$ &$0.13510$  & $0.01842(5)$ & $0.2551(5)$ & $0.0899(2)$\\
&          &$0.13530$  & $0.01597(5)$ & $0.2379(6)$ & $0.0882(2)$\\
&          &$0.13550$  & $0.01353(5)$ & $0.2194(6)$ & $0.0863(2)$\\
&          &$0.13570$  & $0.01109(5)$ & $0.1993(7)$ & $0.0843(2)$\\
&          &$0.13580$  & $0.00985(5)$ & $0.1883(6)$ & $0.0832(2)$\\
\midrule
\multirow{8}{*}{A4} 
&$0.13510$ &$0.13510$  & $0.02557(5)$ & $0.2958(5)$ & $0.0933(2)$\\
&$0.13530$ &$0.13530$  & $0.02061(5)$ & $0.2659(5)$ & $0.0896(2)$\\
&$0.13550$ &$0.13550$  & $0.01571(5)$ & $0.2328(5)$ & $0.0857(2)$\\
&$0.13570$ &$0.13570$  & $0.01087(5)$ & $0.1948(7)$ & $0.0814(2)$\\
&$0.13590$ &$0.13510$  & $0.01574(5)$ & $0.2334(6)$ & $0.0851(2)$\\
&          &$0.13530$  & $0.01330(5)$ & $0.2150(6)$ & $0.0833(2)$\\
&          &$0.13550$  & $0.01087(5)$ & $0.1949(7)$ & $0.0813(2)$\\
&          &$0.13570$  & $0.00845(5)$ & $0.1726(7)$ & $0.0791(2)$\\
&          &$0.13590$  & $0.00601(6)$ & $0.1466(8)$ & $0.0766(3)$\\
\midrule
\multirow{8}{*}{A5} 
&$0.13510$ &$0.13510$  & $0.02488(5)$ & $0.2907(6)$ & $0.0918(2)$\\
&$0.13530$ &$0.13530$  & $0.01993(5)$ & $0.2605(6)$ & $0.0881(2)$\\
&$0.13550$ &$0.13550$  & $0.01504(5)$ & $0.2269(6)$ & $0.0841(2)$\\
&$0.13570$ &$0.13570$  & $0.01021(5)$ & $0.1882(7)$ & $0.0797(2)$\\
&$0.13594$ &$0.13510$  & $0.01460(5)$ & $0.2244(6)$ & $0.0828(2)$\\
&          &$0.13530$  & $0.01217(5)$ & $0.2053(7)$ & $0.0810(2)$\\
&          &$0.13550$  & $0.00975(5)$ & $0.1844(7)$ & $0.0789(2)$\\
&          &$0.13570$  & $0.00734(5)$ & $0.1608(8)$ & $0.0766(3)$\\
&          &$0.13594$  & $0.00444(5)$ & $0.1263(8)$ & $0.0735(3)$\\
\bottomrule
\end{tabular}
\end{center}
\caption{\footnotesize
Partially quenched average quark masses, pseudoscalar meson masses and
decay constants in lattice units for $\beta=5.2$.\label{tab:mpsfps52}}
\end{table}

\begin{table}[p!]
\small
\begin{center}
\begin{tabular}{@{\extracolsep{0.2cm}}lllllll}
\toprule
 &  $\kappa_r$ & $\kappa_s$ & $am_{rs}$ & $am_\mathrm{PS}$ &
 $F_\mathrm{PS}^\mathrm{bare}$  \\
\midrule
\multirow{8}{*}{E4} 
&$0.13540$ &$0.13540$  & $0.03121(7)$ & $0.3018(12)$ & $0.0884(5)$\\
&$0.13560$ &$0.13560$  & $0.02589(8)$ & $0.2746(13)$ & $0.0852(5)$\\
&$0.13580$ &$0.13580$  & $0.02061(8)$ & $0.2449(13)$ & $0.0819(5)$\\
&$0.13600$ &$0.13600$  & $0.01535(8)$ & $0.2118(14)$ & $0.0784(5)$\\
&$0.13610$ &$0.13540$  & $0.02191(8)$ & $0.2529(14)$ & $0.0825(5)$\\
&          &$0.13560$  & $0.01928(8)$ & $0.2372(14)$ & $0.0809(5)$\\
&          &$0.13580$  & $0.01666(8)$ & $0.2206(14)$ & $0.0793(5)$\\
&          &$0.13600$  & $0.01404(9)$ & $0.2028(15)$ & $0.0775(5)$\\
&          &$0.13610$  & $0.01272(9)$ & $0.1934(15)$ & $0.0767(6)$\\
\midrule
\multirow{8}{*}{E5} 
&$0.13540$ &$0.13540$  & $0.02968(3)$ & $0.2896(3)$ & $0.08509(17)$\\
&$0.13560$ &$0.13560$  & $0.02436(3)$ & $0.2619(3)$ & $0.08173(17)$\\
&$0.13580$ &$0.13580$  & $0.01907(3)$ & $0.2318(3)$ & $0.07809(17)$\\
&$0.13600$ &$0.13600$  & $0.01383(3)$ & $0.1979(4)$ & $0.07413(17)$\\
&$0.13625$ &$0.13540$  & $0.01840(3)$ & $0.2281(4)$ & $0.07672(17)$\\
&          &$0.13560$  & $0.01578(3)$ & $0.2114(4)$ & $0.07514(17)$\\
&          &$0.13580$  & $0.01317(3)$ & $0.1934(4)$ & $0.07337(17)$\\
&          &$0.13600$  & $0.01056(3)$ & $0.1737(4)$ & $0.07139(17)$\\
&          &$0.13625$  & $0.00727(3)$ & $0.1454(4)$ & $0.06852(20)$\\
\midrule
\multirow{8}{*}{F6} 
&$0.13540$ &$0.13540$  & $0.02874(2)$ & $0.2824(3)$ & $0.08294(18)$\\
&$0.13560$ &$0.13560$  & $0.02341(2)$ & $0.2542(3)$ & $0.07947(18)$\\
&$0.13580$ &$0.13580$  & $0.01812(3)$ & $0.2234(3)$ & $0.07570(19)$\\
&$0.13600$ &$0.13600$  & $0.01288(3)$ & $0.1885(3)$ & $0.0716(2)$\\
&$0.13635$ &$0.13540$  & $0.01616(3)$ & $0.2113(4)$ & $0.0729(2)$\\
&          &$0.13560$  & $0.01354(3)$ & $0.1935(4)$ & $0.0713(2)$\\
&          &$0.13580$  & $0.01092(3)$ & $0.1741(4)$ & $0.0695(2)$\\
&          &$0.13600$  & $0.00832(3)$ & $0.1524(4)$ & $0.0674(2)$\\
&          &$0.13635$  & $0.00374(3)$ & $0.1036(5)$ & $0.0629(3)$\\
\midrule
\multirow{8}{*}{F7} 
&$0.13540$ &$0.13540$  & $0.028455(17)$ & $0.2801(2)$ & $0.08208(14)$\\
&$0.13560$ &$0.13560$  & $0.023121(19)$ & $0.2520(3)$ & $0.07857(14)$\\
&$0.13580$ &$0.13580$  & $0.017832(19)$ & $0.2211(3)$ & $0.07473(15)$\\
&$0.13600$ &$0.13600$  & $0.012594(20)$ & $0.1860(3)$ & $0.07045(16)$\\
&$0.13638$ &$0.13540$  & $0.01549(2)$ & $0.2068(3)$ & $0.07122(18)$\\
&          &$0.13560$  & $0.01288(2)$ & $0.1886(3)$ & $0.06957(18)$\\
&          &$0.13580$  & $0.01027(2)$ & $0.1687(3)$ & $0.06771(18)$\\
&          &$0.13600$  & $0.00767(2)$ & $0.1463(3)$ & $0.06557(18)$\\
&          &$0.13638$  & $0.00272(2)$ & $0.0886(4)$ & $0.0603(2)$\\
\bottomrule
\vspace{-1cm}
\end{tabular}
\end{center}
\caption{\footnotesize
Partially quenched average quark masses, pseudoscalar meson masses and
decay constants in lattice units for $\beta=5.3$.\label{tab:mpsfps53}}
\end{table}

\begin{table}[p!]
\small
\begin{center}
\begin{tabular}{@{\extracolsep{0.2cm}}lllllll}
\toprule
 &  $\kappa_r$ & $\kappa_s$ & $am_{rs}$ & $am_{\mathrm{PS}}$ &
 $F_\mathrm{PS}^\mathrm{bare}$  \\
\midrule
\multirow{8}{*}{N4} 
&$0.13600$ &$0.13600$  & $0.02308(3)$ & $0.2172(6)$ & $0.0631(4)$\\
&$0.13615$ &$0.13615$  & $0.01885(3)$ & $0.1958(7)$ & $0.0605(4)$\\
&$0.13630$ &$0.13630$  & $0.01462(3)$ & $0.1723(8)$ & $0.0578(4)$\\
&$0.13645$ &$0.13645$  & $0.01040(3)$ & $0.1456(8)$ & $0.0548(4)$\\
&$0.13650$ &$0.13600$  & $0.01602(3)$ & $0.1805(8)$ & $0.0584(4)$\\
&          &$0.13615$  & $0.01391(3)$ & $0.1682(7)$ & $0.0572(4)$\\
&          &$0.13630$  & $0.01181(3)$ & $0.1550(8)$ & $0.0558(4)$\\
&          &$0.13645$  & $0.00970(3)$ & $0.1407(9)$ & $0.0543(4)$\\
&          &$0.13650$  & $0.00899(3)$ & $0.1357(8)$ & $0.0538(4)$\\
\midrule
\multirow{8}{*}{N5} 
&$0.13600$ &$0.13600$  & $0.02265(3)$ & $0.2144(7)$ & $0.0622(4)$\\
&$0.13615$ &$0.13615$  & $0.01841(3)$ & $0.1927(7)$ & $0.0595(4)$\\
&$0.13630$ &$0.13630$  & $0.01419(3)$ & $0.1689(7)$ & $0.0566(4)$\\
&$0.13645$ &$0.13645$  & $0.00997(4)$ & $0.1417(8)$ & $0.0534(4)$\\
&$0.13660$ &$0.13600$  & $0.01417(4)$ & $0.1689(7)$ & $0.0558(4)$\\
&          &$0.13615$  & $0.01207(4)$ & $0.1559(7)$ & $0.0546(3)$\\
&          &$0.13630$  & $0.00996(4)$ & $0.1418(7)$ & $0.0532(3)$\\
&          &$0.13645$  & $0.00786(4)$ & $0.1262(8)$ & $0.0516(3)$\\
&          &$0.13660$  & $0.00575(4)$ & $0.1085(8)$ & $0.0499(4)$\\
\midrule
\multirow{8}{*}{N6} 
&$0.13600$ &$0.13600$  & $0.022299(17)$ & $0.2099(6)$ & $0.0609(3)$\\
&$0.13615$ &$0.13615$  & $0.018062(17)$ & $0.1881(6)$ & $0.0581(3)$\\
&$0.13630$ &$0.13630$  & $0.013835(18)$ & $0.1642(6)$ & $0.0552(2)$\\
&$0.13645$ &$0.13645$  & $0.009619(19)$ & $0.1369(6)$ & $0.0519(3)$\\
&$0.13667$ &$0.13600$  & $0.01283(2)$ & $0.1586(6)$ & $0.0534(3)$\\
&          &$0.13615$  & $0.01073(2)$ & $0.1450(5)$ & $0.0522(3)$\\
&          &$0.13630$  & $0.00863(2)$ & $0.1301(5)$ & $0.0507(3)$\\
&          &$0.13645$  & $0.00653(2)$ & $0.1135(5)$ & $0.0491(3)$\\
&          &$0.13667$  & $0.00343(3)$ & $0.0834(7)$ & $0.0461(3)$\\
\midrule
\multirow{8}{*}{O7} 
&$0.13600$ &$0.13600$  & $0.022117(10)$ & $0.2085(3)$ & $0.06040(14)$\\
&$0.13620$ &$0.13620$  & $0.016468(10)$ & $0.1789(3)$ & $0.05666(14)$\\
&$0.13640$ &$0.13640$  & $0.010837(12)$ & $0.1447(3)$ & $0.05241(14)$\\
&$0.13660$ &$0.13660$  & $0.005226(14)$ & $0.1012(3)$ & $0.04741(15)$\\
&$0.13671$ &$0.13600$  & $0.012084(13)$ & $0.1535(3)$ & $0.05225(18)$\\
&          &$0.13620$  & $0.009283(13)$ & $0.1345(3)$ & $0.05050(16)$\\
&          &$0.13640$  & $0.006483(13)$ & $0.1126(3)$ & $0.04841(16)$\\
&          &$0.13660$  & $0.003681(15)$ & $0.0855(3)$ & $0.04581(16)$\\
&          &$0.13671$  & $0.002130(15)$ & $0.0658(4)$ & $0.0440(2)$\\
\bottomrule
\vspace{-1cm}
\end{tabular}
\end{center}
\caption{\footnotesize
Partially quenched average quark masses, pseudoscalar meson masses and
decay constants in lattice units for $\beta=5.5$. 
\label{tab:mpsfps55}}
\end{table}

\end{appendix}
\newpage

\providecommand{\href}[2]{#2}\begingroup\raggedright\endgroup
\end{document}